\documentclass[usenatbib]{mn2e}
\usepackage{natbibmnfix,graphicx,times}

\newcommand{\cmsq}{\mbox{ cm$^2$}}

\newcommand{\eV}{\mbox{ eV}}
\newcommand{\kel}{\mbox{ K}}

\newcommand{\Mpc}{\mbox{ Mpc}}

\newcommand{\Msun}{\mbox{ M$_\odot$}}

\newcommand{\hunits}{\mbox{ km s$^{-1}$ Mpc$^{-1}$}}
\newcommand{\kms}{\mbox{ km s$^{-1}$}}
\newcommand{\bq}{\begin{equation}}
\newcommand{\eq}{\end{equation}}
\newcommand{\bqa}{\begin{eqnarray}}
\newcommand{\eqa}{\end{eqnarray}}
\newcommand{\deriv}{{\rm d}}

\newcommand{\bxio}{\bar{x}_i}
\newcommand{\bxh}{\bar{x}_{\rm HI}}

\newcommand{\hii}{HII }
\newcommand{\lya}{Ly$\alpha$ }

\newcommand{\lyg}{Ly$\gamma$ }
\newcommand{\recunits}{\mbox{ cm$^{3}$ s$^{-1}$}}

\newcommand{\apj}{ApJ}
\newcommand{\apjl}{ApJ}
\newcommand{\apjs}{ApJS}

\newcommand{\aj}{AJ}
\newcommand{\mnras}{MNRAS}
\newcommand{\physrep}{Physics Reports}
\newcommand{\nat}{Nature}

\title[Recombinations and Reionization]{Taxing the rich:  recombinations and bubble growth during reionization}

\author[Furlanetto \& Oh]{Steven R. Furlanetto$^1$\thanks{Email:sfurlane@tapir.caltech.edu} and S. Peng Oh$^2$\thanks{Email: peng@physics.ucsb.edu} \\ 
$^1$Division of Physics, Mathematics, \& Astronomy; California Institute of Technology; Mail Code 130-33; Pasadena, CA 91125 \\ $^2$Department of Physics; University of California; Santa Barbara, CA 93106} 

\begin{document}

\maketitle

\begin{abstract}
Reionization is inhomogeneous for two reasons: the clumpiness of the intergalactic medium (IGM) and clustering of the discrete ionizing sources. While numerical simulations can in principle take both into account, they are at present limited by small box sizes. On the other hand, analytic models have only examined the limiting cases of a clumpy IGM (with uniform ionizing emissivity) and clustered sources (embedded in a uniform IGM). Here, we present the first analytic model that includes both factors.  At first, recombinations can be ignored and ionized bubbles grow primarily through major mergers, because at any given moment the bubbles have a well-defined characteristic size.  As a result, reionization resembles ``punctuated equilibrium," with a series of well-separated sharp jumps in the ionizing background.  These features are local effects and do not reflect similar jumps in the global ionized fraction.  We then combine our bubble model with a simple description of recombinations in the IGM.  We show that the bubbles grow until recombinations balance ionizations, when their expansion abruptly halts.  If the IGM density structure is similar to that at moderate redshifts, this limits the bubble radii to $\sim 20$ comoving Mpc; however, if the IGM is significantly clumpier at higher redshifts (because of minihalo formation, for example), the limit could be much smaller.  Once a bubble reaches saturation, that region of the universe has for all intents and purposes entered the ``post-overlap" stage.  Because different \hii regions saturate over a finite time interval, the overlap epoch actually has a finite width.  Our model also predicts a mean recombination rate several times larger than expected for a uniformly-illuminated IGM.  This picture naturally explains the substantial large-scale variation in Lyman-series opacity along the lines of sight to the known $z>6$ quasars.  More quasar spectra will shed light on the transition between the ``bubble-dominated" topology characteristic of reionization and the ``web-dominated" topology characteristic of the later universe.

\end{abstract}


\section{Introduction}
\label{intro}

The reionization of hydrogen in the intergalactic medium (IGM) is a landmark event in the history of structure formation:  it is the epoch at which (radiative) feedback from the first generation of sources impacts every baryon in the universe, and thus the first time that galaxies can be said to influence the course of structure formation throughout the IGM.  As such, a good deal of observational and theoretical effort has gone into understanding this era.  The first -- and simplest -- question to answer is when reionization occurred.  Currently, the observations paint a complex picture.  The Sloan Digital Sky Survey (SDSS) has discovered several $z>6$ quasars. Each shows saturated or nearly saturated absorption in the Ly$\alpha$ forest as well as a rapid decrease in the transmission at $z \approx 6$ \citep{fan01,becker01,white03,fan03,fan04}.  The appearance of the long-sought \citet{gunn65} absorption trough is a crucial constraint on the ionizing background of the universe.  Although the optical depth only limits the global neutral fraction to be $\bxh \ga 10^{-3}$ \citep{fan02,songaila04,oh05}, its rapid evolution may be characteristic of the tail end of reionization \citep{gnedin00,razoumov02}.  On the other hand, measurements of the polarization of the cosmic microwave background (CMB) by the \emph{Wilkinson Microwave Anisotropy Probe} (\emph{WMAP}) found a high optical depth to electron scattering for CMB photons, which requires reionization to begin at $z \ga 15$ \citep{kogut03}, albeit with large uncertainties.

A number of other observations provide additional constraints, though their interpretation is even more difficult.  These include: the IGM temperature at $z \la 4$ \citep{theuns02-reion,hui03}, which suggests relatively late reionization; detailed modeling of the proximity zones around $z>6$ quasars \citep{wyithe04-prox,mesinger04,wyithe05-prox}, which suggests $\bxh \ga 0.1$ at $z \approx 6.3$; and the lack of evolution in the luminosity function of Ly$\alpha$-selected galaxies over $z \approx 5.7$--$6.5$ \citep{malhotra04}, which suggests reionization at an earlier epoch (although this is subject to uncertainties about the clustering of high-redshift galaxies; \citealt{furl04-lya,wyithe05-bub}).

Weaving these individual threads into a consistent picture requires a rather complex history.  In the simplest models, reionization proceeds extremely rapidly because galaxy formation is nearly exponential at high redshifts (e.g., \citealt{barkana01} and references therein).  This is incompatible with the combination of SDSS and \emph{WMAP} measurements (at least if we ignore the large errors on the latter measurement).  The most popular solution is to invoke some sort of feedback  mechanism to decrease the ionizing efficiency of galaxies during reionization (e.g., \citealt{wyithe03,wyithe03-letter,cen03-letter,cen03,haiman03}).  Feedback may directly affect collapsed objects (for example, through enrichment by heavy elements) or the IGM itself (for example, through photoheating).  Unfortunately, the underlying mechanism -- or mechanisms -- remain mysterious, as do their effects on the reionization history (though they are unlikely to introduce sharp features; \citealt{furl05-double}).  Obviously, the physics of reionization, the first luminous sources, and the IGM are deeply intertwined.  Disentangling them requires new methods to study the high-redshift universe.

Fortunately, the existing quasar data hints at one promising direction:  the two best-studied lines of sight contain remarkable differences \citep{becker01,white03}.  SDSS J1030+0524, at $z=6.28$, shows completely saturated absorption from $z \approx 5.97$--$6.18$ in all the Lyman transitions.  This places a lower limit on the (mean) \lya optical depth $\tau_{\alpha} > 9.9$ (at the $2\sigma$ level).  In contrast, SDSS J1148+5251, at $z=6.42$, shows a single transmission feature in Ly$\alpha$ at $z=6.08$, a number of obvious transmission gaps in Ly$\beta$ over $z=5.95$--$6.18$, and a faint continuum in Ly$\alpha$, Ly$\beta$, and Ly$\gamma$ from $z=6.18$--$6.33$.  While at first attributed to contamination by a $z=5$ interloper, \citet{oh05} and \citet{white05} recently demonstrated that both the continuum and the spikes are true transmission from the $z \sim 6$ IGM.  \citet{oh05} used the \lyg trough to place an upper bound on the effective \lya optical depth at $z=6.2$--$6.3$ of $\tau_{\alpha} < 15.4$ ($2\sigma$), with the likely value in the range $\tau_{\alpha} \sim 7$--$11$ (the large uncertainty comes primarily from the unknown IGM density distribution).  At lower redshifts, the clear transmission spikes toward SDSS J1148+5251 imply even more significant differences in the IGM transmission properties between these lines of sight.  Both therefore show large jumps in the ionizing background at $z \sim 6$ but are otherwise entirely different.  Most interestingly, the differences are not ``one-shot" affairs that can be localized to a small portion of the spectrum:  instead they cohere over $\sim 100$ comoving Mpc scales.  How can we explain modulation in the transmission over such large scales, on which density fluctuations should be tiny at $z \sim 6$?  And how can we reconcile it with the sudden and rapid evolution of the mean transmission along each line of sight?

To answer these questions, we must understand how the spatial distribution of ionized gas evolves through reionization.  Early models, such as \citet{arons72}, followed the growth of \hii regions around individual galaxies.  \citet[hereafter FZH04]{furl04-bub} showed that the sizes are actually determined by clustering of the underlying galaxy population.  Thus studying the inhomogeneity inherent to the reionization process can open new windows on the physics of reionization. FZH04 showed that the ionized bubbles are large throughout reionization and that they have a characteristic size.  This offers a first step in explaining how large-scale fluctuations develop.  We review their model in \S \ref{fzh}.  In this paper, we focus on how the \hii regions evolve during reionization and especially near its completion.  In \S \ref{eps}, we examine how the bubbles actually grow and merge with each other in order to elucidate the meanings of jumps in the ionizing background.  We show that ``major mergers" dominate the growth process.  As a result, the radiation background at any point in the IGM evolves through a series of discrete and rapid jumps.   However, unlike in previous theoretical treatments (e.g., \citealt{gnedin00}), these jumps do \emph{not} correspond to global overlap; rather, they are simply the byproduct of inhomogeneous reionization.

In the original model of FZH04, the bubbles merge until they form a single infinitely large object at the completion of reionization.  However, by ignoring IGM substructure the FZH04 model treats recombinations extremely crudely.  \citet[hereafter MHR00]{miralda00} presented a complementary model for inhomogeneous reionization that focuses precisely on the question of recombinations.  They argued that locally reionization proceeds from low to high-density regions.\footnote{This stands in contrast to the large scale behavior, which traces clusters of sources inside overdense regions.}  Once ionized, voids remain so because of their long recombination times, but dense clumps (corresponding to filaments or bound structures) rapidly return to neutrality.  Thus, within an \hii region, the ionized fraction traces the fluctuating density field, with the neutral gas confined to smaller and denser clumps as reionization proceeds.  We describe this model in more detail in \S \ref{density}, and we examine some of the subtleties of the MHR00 treatment in an Appendix.

In a sense, FZH04 and MHR00 were designed to address different regimes. The former applies best during the pre-overlap stage, when the discrete bubble structure from source clustering is most important, while the latter applies best during the post-overlap stage, when the ionizing background is nearly uniform and the local ionized fraction traces IGM density fluctuations. Of course, toward the tail end of reionization, both contributions are equally important. We thus show how to integrate the FZH04 and MHR00 models in \S \ref{recomb}. The large ionized bubbles of the FZH04 model are of course not fully ionized; rather, they contain dense neutral clumps that do not affect the bubble topology so long as the mean free path between clumps exceeds the bubble radius.  As a bubble grows, the ionized gas must therefore extend deeper into the dense clumps, and the recombination rate increases.  Eventually, these internal recombinations consume all the ionizing photons, and the bubble stops growing.  For the bubble interior, this point is effectively the end of ``overlap" because bubble mergers have no further effect on the ionizing background.  Thus overlap is actually a local phenomenon.  In \S \ref{obs}, we show that the resulting saturation has a significant impact on the Lyman-series transmission along quasar lines of sight toward the tail end of reionization.   Finally, we conclude in \S \ref{disc}.

In our numerical calculations, we assume a cosmology with $\Omega_m=0.3$, $\Omega_\Lambda=0.7$, $\Omega_b=0.046$, $H=100 h \hunits$ (with $h=0.7$), $n=1$, and $\sigma_8=0.9$, consistent with the most recent measurements \citep{spergel03}.  Unless otherwise specified, we use comoving units for all distances.

\begin{figure*}
\begin{center}
\resizebox{8cm}{!}{\includegraphics{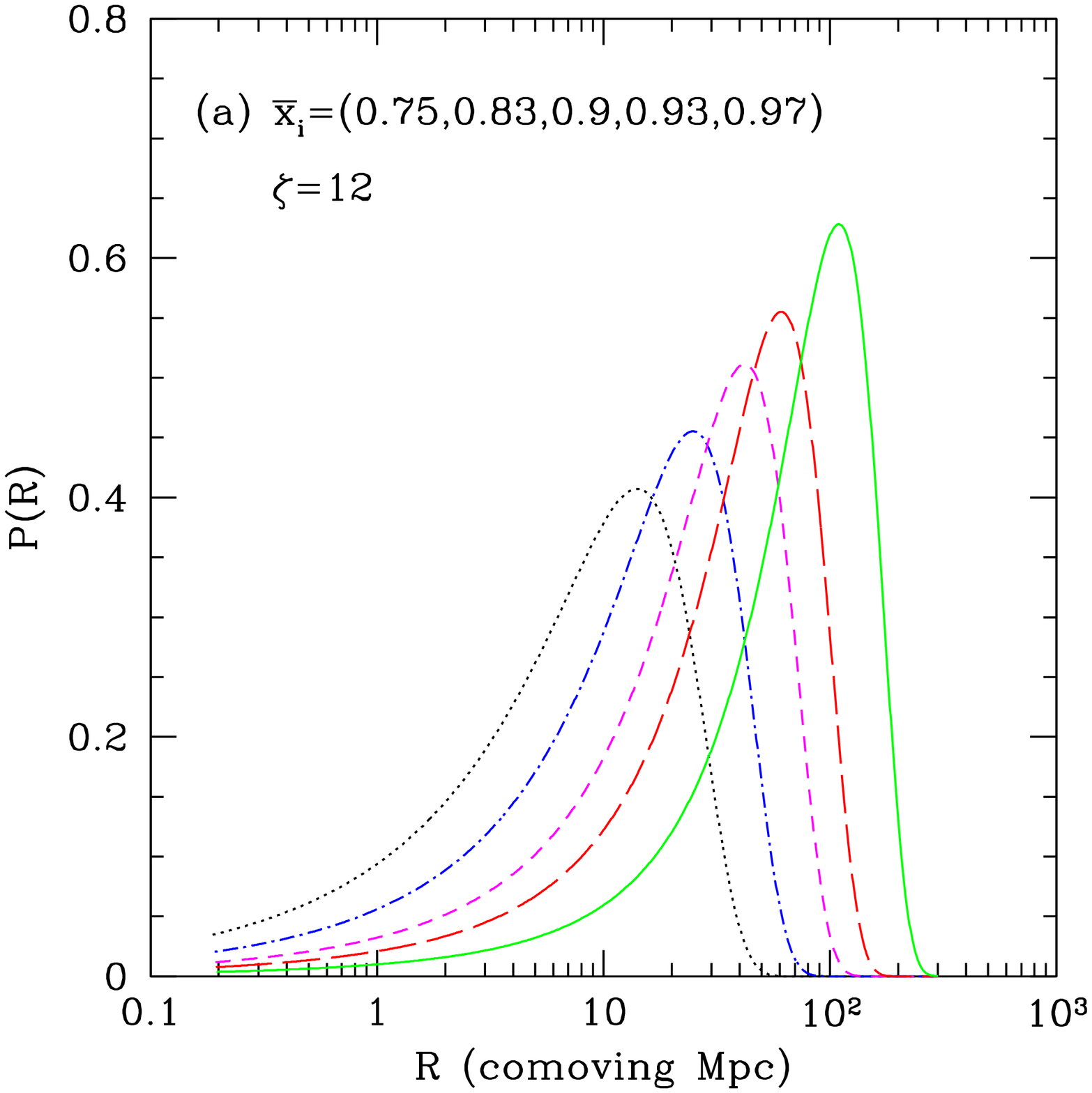}}
\hspace{0.13cm}
\resizebox{8cm}{!}{\includegraphics{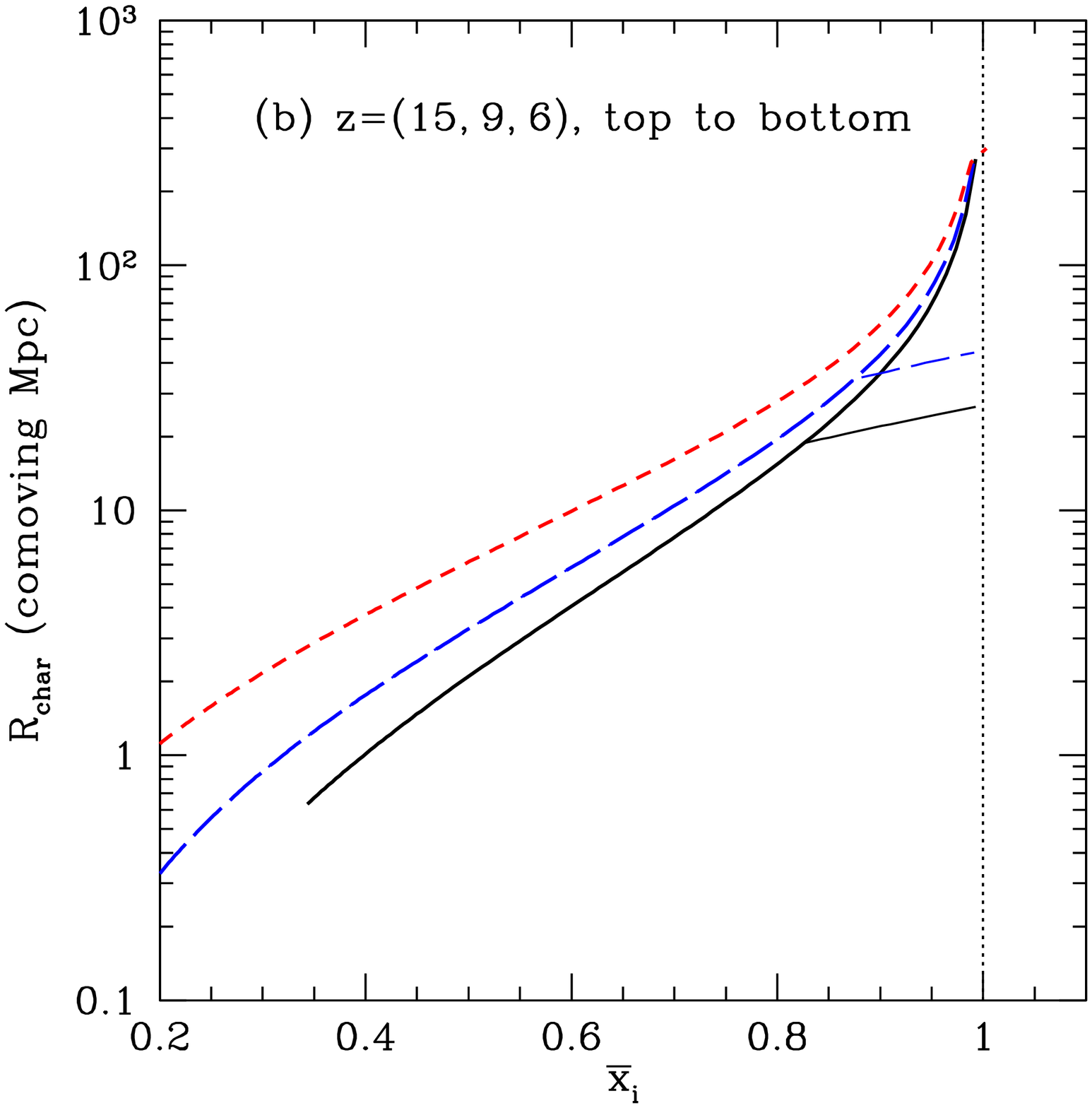}}
\end{center}
\caption{\emph{(a)}: Probability distribution of mean free paths \emph{in ionized regions}.  The different curves correspond to different $\bar{x}_i$.  All curves take $\zeta=12$ and range from $z=8.8$--$8$; we do not take recombinations into account.  \emph{(b)}:   Evolution of characteristic bubble size (or mean free path) with $\bar{x}_i$ at a fixed redshift (which corresponds to varying $\zeta$).  The solid, long-dashed, and short-dashed curves are for $z=6,\,9$, and $15$, respectively. The thick lines assume the FZH04 model; the thin lines (shown only for $z=6$ and 9) incorporate recombinations as described in \S \ref{recomb}.}
\label{fig:pmfp}
\end{figure*}

\section{The Bubble Size Distribution Without Recombinations}
\label{fzh}

Recent numerical simulations (e.g., \citealt{ciardi03-sim,sokasian03,sokasian04}) show that reionization proceeds ``inside-out'' from high density clusters of sources to voids (at least in the standard scenario where galaxies ionize the IGM).  We therefore associate HII regions with large-scale overdensities.  We assume that a galaxy of mass $m_{\rm gal}$ can ionize a mass $\zeta m_{\rm gal}$, where $\zeta$ is a constant that depends on (among other things) the efficiency of ionizing photon production, the escape fraction of these photons from the host galaxy, the star formation efficiency, and the mean number of recombinations.  Each of these quantities is highly uncertain (e.g., massive Population III stars can dramatically increase the ionizing efficiency; \citealt{bromm01-vms}), but at least to a rough approximation they can be collapsed into this single efficiency factor.  The criterion for a region to be ionized by the galaxies contained inside it is then $f_{\rm coll} > \zeta^{-1}$, where $f_{\rm coll}$ is the fraction of mass bound in halos above some $m_{\rm min}$.  We will assume that this minimum mass corresponds to a virial temperature $T_{\rm vir}=10^4 \kel$, at which point hydrogen line cooling becomes efficient.  In the extended Press-Schechter model \citep{lacey93}, this places a condition on the mean overdensity within a region of mass $m$,
\bq 
\delta_m \ge \delta_x(m,z) \equiv
\delta_c(z) - \sqrt{2} K(\zeta) [\sigma^2_{\rm min} -
\sigma^2(m)]^{1/2},
\label{eq:deltax}
\eq
where $K(\zeta) \equiv {\rm erf}^{-1}(1 - \zeta^{-1})$, $\sigma^2(m)$ is the variance of density fluctuations on the scale $m$, $\sigma^2_{\rm min} \equiv \sigma^2(m_{\rm min})$, and $\delta_c(z)$ is the critical density for collapse.
The {\it global} ionized fraction is $\bxio=\zeta f_{\rm coll,g}$, where $f_{\rm coll,g}$ is the global collapse fraction.

FZH04 showed how to construct the mass function of HII regions from $\delta_x$ in an analogous way to the halo mass function \citep{press74,bond91}.  The barrier in equation (\ref{eq:deltax}) is well approximated by a linear function in $\sigma^2$, $\delta_x \approx B(m,z) \equiv B_0 + B_1 \sigma^2(m)$. In that case the mass function of ionized bubbles (i.e., the comoving number density of \hii regions with masses in the range $m \pm \deriv m/2$) has an analytic solution \citep{sheth98,mcquinn05}:
\bq
n_b(m,z) = \sqrt{\frac{2}{\pi}} \ \frac{\bar{\rho}}{m^2} \ \left|
  \frac{\deriv \ln \sigma}{\deriv \ln m} \right| \ \frac{B_0(z)}{\sigma(m)} \exp \left[ - \frac{B^2(m,z)}{2 \sigma^2(m)} \right],
\label{eq:dndm}
\eq
where $\bar\rho$ is the mean density of the universe.  FZH04 showed some examples of how the bubble sizes evolve throughout the early and middle stages of reionization (see also Fig.~\ref{fig:merger}\emph{a}).  Figure~\ref{fig:pmfp}\emph{a} plots the distributions near the end of reionization (the era with which we will be most concerned) assuming $\zeta=12$.  We have phrased the size distribution in terms of the mean free path of ionizing photons by equating that quantity to the bubble radius:  this implicitly neglects absorption within the bubbles, whose importance we consider in \S \ref{recomb} below.  Panel \emph{a} shows the probability distribution $P(R) \deriv R$ of the mean free path, or the fraction of ionized matter within bubbles of comoving size $R \pm \deriv R/2$.  

Figure~\ref{fig:pmfp}\emph{a} illustrates a number of key points about the model.  First, the bubbles are large:  tens of comoving Mpc during the final stages.  This is because high-redshift galaxies cluster strongly, so small mean overdensities translate to much larger overdensities of galaxies.  Furthermore, the bubbles attain a well-defined characteristic size $R_{\rm char}$ at any point during reionization.  This occurs because of the Gaussian nature of the underlying linear density fluctuations and because the barrier $\delta_x$ is a (decreasing) function of $m$.  Near the completion of reionization, over- and underdense regions sit in large and small bubbles, respectively.  In contrast, the barrier used in constructing the halo mass function, $\delta_c(z)$, is independent of mass, which yields the usual power law behavior at small masses.  

The thick curves in Figure~\ref{fig:pmfp}\emph{b} show how this characteristic size depends on $\bxio$ and redshift.  We vary $\zeta$ to fix $\bxio$ at $z=6,\,9$, and $15$ (solid, long-dashed, and short-dashed curves, respectively).  Regardless of redshift, $R_{\rm char}$ surpasses $1 \Mpc$ relatively early in reionization (note that this is already much larger than an \hii region around an individual galaxy) and then increases at an accelerating rate.  Bubbles become larger if reionization happens earlier, because the underlying galaxy population is more biased at high redshifts.  However, by the late stages ($\bxio \ga 0.75$), the differences are relatively small.  For our purposes, the crucial result is that $R_{\rm char}$ rapidly approaches infinity when $\bxio \ga 0.9$.

As emphasized by FZH04, this simple analytic model makes a number of assumptions about the reionization process.  But the most glaring is that the bubbles are roughly spherical:  clearly the inhomogeneous distribution of sources and the cosmic web will introduce asymmetries throughout reionization.  This is not as bad as it may sound, because the model actually computes the total mass of ionized regions; the radius is therefore only a characteristic value, and it may still be relevant.  Simulations like \citet{sokasian03} and \citet{ciardi03-sim} do seem to produce roughly spherical regions throughout most of reionization.  However, a more serious problem is that the topology may reverse itself late in reionization:  the universe could contains neutral islands embedded in an ionized sea, rather than the other way around.  This certainly happens in simulations, although the timing is suspect because of their small sizes.  In the FZH04 model, the universe is instead divided into isolated bubbles separated by thin walls of mostly neutral gas.  That is clearly an approximation, but it still may be true that $R_{\rm char}$ gives the typical distance to a neutral island.  The regime over which the model applies must be determined through detailed comparison to numerical simulations; however, simulations on the relevant scales are not yet available (and the size distribution of \hii regions has yet to be quantified in any existing simulations!).  Another crucial assumption is that the recombination rate can be approximated as uniform across the entire universe:  the purpose of \S \ref{recomb} is to test the validity of this assumption.  Along the way we will see that it also alleviates the topological problem, and that $R_{\rm char}$ describes the typical mean free path of an ionizing photon. 

\section{Bubble Growth in the Excursion Set Model}
\label{eps}

Figure~\ref{fig:pmfp}\emph{b} shows that bubbles grow rapidly, especially in the late stages of reionization.  In this section we will study the mechanisms driving that evolution.  

\subsection{The progenitor size distribution}
\label{prog}

One advantage of the excursion set formulation of the dark matter halo mass function is that conditional mass functions follow easily \citep{lacey93}.  The same holds true for the bubble mass function.  Suppose we consider a region of the universe with overdensity $\delta_M$ on scale $M_b$ corresponding to $\sigma^2_b$.  The derivation of the mass function follows exactly the same procedure as in FZH04, except that the random walk begins at the point $(\delta_M,\sigma^2_b)$ so the (linear) absorbing barrier has a different amplitude.  In particular, we can imagine that our initial point corresponds to a bubble of size $M_b$ at $z_b$:  in that case, $\delta_M=B(M_b,z_b)$, and the conditional mass function represents the bubble progenitor distribution at some $z>z_b$.  It is easy to show that
\bqa
n_b(m,z|M_b,z_b) & = & \sqrt{\frac{2}{\pi}} \  \frac{\bar{\rho}}{m^2} \ \left|
  \frac{\deriv \ln \sigma}{\deriv \ln m} \right| \nonumber \\
  & & \times \frac{\sigma^2[B(M_b,z)-B(M_b,z_b)]}{(\sigma^2-\sigma_b^2)^{3/2}} 
\nonumber \\ & & \times \exp \left\{ - \frac{[B(m,z)-B(M_b,z_b)]^2}{2 (\sigma^2-\sigma_b^2)} \right\}.
\label{eq:prog}
\eqa
Note that this has the same structure as the unconditional $n_b(m)$; the only differences are the substitutions
\bqa
B_0(m,z) & \rightarrow & B(M_b,z) - B(M_b,z_b) \nonumber \\
B(m,z) & \rightarrow & B(m,z) - B(M_b,z_b) \\
\sigma^2 & \rightarrow & \sigma^2 - \sigma_b^2, \nonumber
\label{eq:condsubs}
\eqa
which simply represent translating the origin in the diffusion problem.  It is also structurally similar to the halo progenitor distribution \citep{lacey93}, with the exception of a linearly-increasing barrier, of course.  

\begin{figure}
\begin{center}
\resizebox{8cm}{!}{\includegraphics{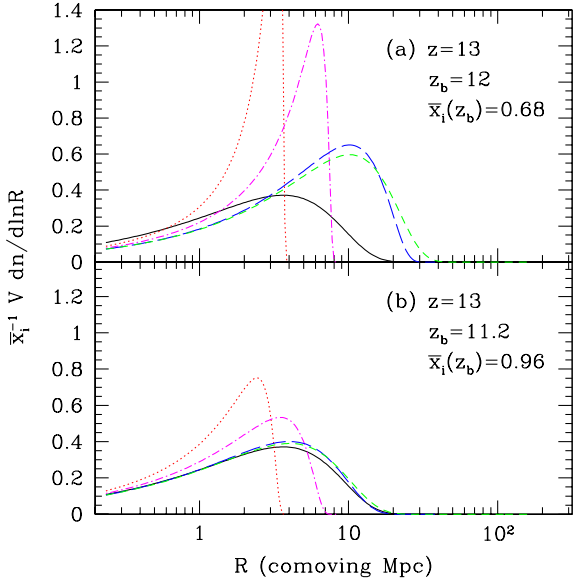}}\\%
\end{center}
\caption{Progenitor bubble distributions.  \emph{(a)}:  The solid curve shows the overall size distribution at $z=13$ if $\zeta=40$ (corresponding to $\bxio=0.43$).  The other curves show the size distribution at $z=13$ inside an ionized bubble that at 
$z_b=12$ ($\bxio=0.68$) has $M_b=10^{13},\, 10^{14},\, 10^{16}$, and $10^{18} \Msun$ (dotted, dot-dashed, long-dashed, and short-dashed curves, respectively).  \emph{(b)}:  Same, except $z_b=11.2$ ($\bxio=0.96$).}
\label{fig:progfig}
\end{figure}

Figure~\ref{fig:progfig} shows some example distributions at $z=13$ assuming $\zeta=40$.  Each panel shows the fraction of the universe in bubbles of a given radius, normalized by the mean ionized fraction $\bxio=0.43$.  The solid curve shows the unconditional $n_b(m,z)$.  The dotted, dot-dashed, long-dashed, and short-dashed curves show $n_b(m,z|M_b,z_b)$ if $z_b=12$ and $M_b=10^{13},\,
  10^{14},\, 10^{16}$, and $10^{18} \Msun$; these correspond to $R_b=3.9,\, 8.4,\, 38.9$ and $180 \Mpc$. Panels \emph{a} and \emph{b} assume $z_b=12$ (when $\bxio=0.68$) and $z_b=11.2$ ($\bxio=0.96$), respectively.  Note that the areas under these curves are \emph{not} normalized to unity, because the local ionized fraction at $z=13$ is greater than the mean value.  There are a number of interesting features to the progenitor distribution.  First, large ionized regions tend to contain larger-than-average bubbles at earlier times.  (This is not true for small bubbles, because progenitors cannot be larger than the final object.)  This is of course because such bubbles encompass overdense regions where structure formation begins earlier than average, so ``local overlap" occurs at an earlier time.  However, one might naively expect this bias to become ever stronger as the final bubble size increases.  In fact the conditional $n_b$ approaches a constant form, because $\sigma_b^2 \rightarrow 0$ and $B(M_b,z_b) \rightarrow B_0(z_b)$ on large scales, so equation (\ref{eq:prog}) approaches a constant functional form.  Physically, mass fluctuations on such large scales are extremely small.  Thus, all large bubbles have essentially the same underlying source population and hence essentially the same $n_b$.  It is only at the end of reionization, when density fluctuations appear amplified by the proximity to the phase transition at full ionization, that such low-amplitude large-scale modes contribute to differences in the bubble population.  By that time even a small number of extra sources can connect the already-large bubbles into one fell beast.  Earlier on, this tiny bias does not manifest itself as clearly because the extra sources only ionize a small physical volume.

As a consequence, the progenitor distribution remains relatively unbiased if $z_b$ is near full reionization.  Mathematically, this occurs because $B(M_b,z_b) \approx 0$.  Physically, the mean overdensity in ionized regions is extremely small and the boost in the number density of ionizing sources is modest -- even if the linear bias is large; thus their progenitors are nearly average parts of the universe.  This can also be seen in Figure~\ref{fig:pmfp}, which shows how the size distribution becomes narrower as $\bxio \rightarrow 1$.  Thus, the progenitor bias shows up most strongly early in reionization.  These conclusions are independent of our choice for $\zeta$.

\begin{figure}
\begin{center}
\resizebox{8cm}{!}{\includegraphics{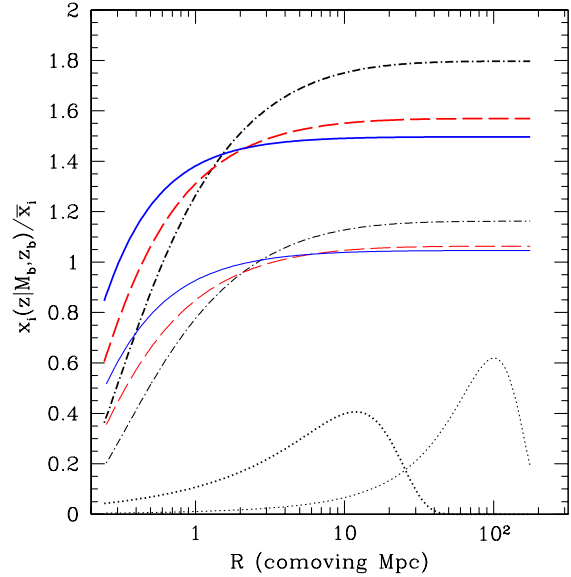}}\\%
\end{center}
\caption{History of the ionized fraction within bubbles relative to the global mean.  The upper (thick) and lower (thin) sets of curves take $z_b=12$ and $11.2$; the ordinate $R$ refers to the final bubble size at $z_{b}$.   Dot-dashed, dashed, and solid curves are for $z=18,\, 15$, and $13$.  The thick and thin dotted curves show the normalized bubble size distributions at $z_b=12$ and $11.2$, respectively.  All assume $\zeta=40$.}
\label{fig:qb}
\end{figure}

Another way of looking at the progenitor bias is to consider the evolution of the ionized fraction within bubbles, which is simply proportional to their collapse fraction.  In the linear barrier approximation,
\bq
x_i(z|M_b,z_b) = \zeta \ {\rm erfc} \left[ \frac{\delta_c(z) - B(M_b,z_b)}{\sqrt{2(\sigma_{\rm min}^2-\sigma_b^2)}} \right]
\label{eq:qcond}
\eq
The upper sets of curves in Figure~\ref{fig:qb} show $x_i(z|M_b,z_b)/\bxio(z)$ at several redshifts if $\zeta=40$ and $z_b=12$ (thick curves) and $11.2$ (thin curves).  The dot-dashed, dashed, and solid curves are for $z=18,\, 15$, and $13$, respectively; these have $\bxio=0.034,\, 0.17$, and $0.43$.  For reference, the thick and thin dotted curves in Figure~\ref{fig:qb} show the bubble size distributions at $z_b=12$ and $11.2$.  Obviously, large bubbles tend to be ionized earlier than average, while small bubbles tend to be ionized later.  The latter occurs because bubbles much smaller than the characteristic size appear only in underdense regions, which contain few ionizing sources.  At large masses, the ratio approaches a constant.  This is clear from equation (\ref{eq:qcond}), because 
$B(M_b) \rightarrow B_0$ and $\sigma_b \ll \sigma_{\rm min}$ for $M_b \rightarrow \infty$.  Physically, the explanation is the same as for the progenitor distribution:  large-scale modes only add a small number of sources that cannot ionize much of the IGM but can connect existing \hii regions.  Thus these modes only affect the topology when $\bxio \approx 1$.  For the same reason, we also see that the bias increases if we select bubbles early on:  when $\bxio \sim 1$, ionized regions have nearly the mean density and hence their histories are near average as well.  On the other hand, for a given bubble the ionization bias is largest early in reionization when galaxies are the most biased, because high-redshift galaxies are so far off on the nonlinear-mass tail that a small boost to the argument of the error function increases $\bxio$ by a large amount.  

We have assumed throughout that $\sigma_{\rm min}$ is fixed at a mass scale corresponding to $T_{\rm vir} = 10^{4}$K and that $\zeta$ is independent of mass. In reality, a variety of feedback effects may decrease the star formation efficiency in shallow potential wells, effectively increasing $m_{\rm min}$ or even turning $\zeta$ into an increasing function of mass. This would only exacerbate the ionization bias discussed above, so long as we normalize to a fixed $\bar{x}_{i}(z)$, because it would increase the  mean bias of the sources. 

\subsection{Merger rates}
\label{merger}

We can use the above expressions to derive approximate merger rates for ionized bubbles.  The procedure is exactly the same as in \citet{lacey93}; the result is
\bqa
\frac{\deriv^2p(m_1,m_T,t)}{\deriv m_2 \ \deriv t} & = & \sqrt{ \frac{2}{\pi}} \ \frac{1}{t} \ \left| \frac{\deriv \ln B}{\deriv \ln t} \right| \ \left| \frac{\deriv \ln \sigma_T}{\deriv \ln m_T} \right| \nonumber \\ & & \times \left( \frac{1}{m_T} \right) \ \frac{B(m_T,z)}{\sigma_T(1-\sigma_T^2/\sigma_1^2)^{3/2}} \nonumber \\ 
& & \times \exp \left[ - \frac{B_0^2(z)}{2} \left( \frac{1}{\sigma_T^2} - \frac{1}{\sigma_1^2} \right) \right],
\label{eq:mergerate}
\eqa
where $\deriv^2p(m_1,m_T,t)/\deriv m_2 \, \deriv t$ is the probability that a given halo of mass $m_1$ will merge with a halo of mass $m_2 \equiv m_T-m_1$ per unit time.  We will also define a merger kernel
\bqa
Q(m_1,m_2,t) \equiv \frac{1}{n_b(m_2,t)} \ \frac{\deriv^2p(m_1, m_T,t)}{\deriv m_2 \ \deriv t}
\label{eq:mergekernel}
\eqa
so that the total rate at which bubbles of mass $m_1$ merge with those of mass $m_2$ per unit volume is $n_b(m_1) n_b(m_2) Q(m_1,m_2)$.

Before proceeding further, we note that this expression suffers from the same limitations as the usual extended Press-Schechter merger rates (see \citealt{benson05} for a detailed discussion).  The most obvious shortcoming is that $Q(m_1,m_2)$ is not symmetric in its arguments, which becomes especially important for large mass ratio mergers.  Moreover, the merger rates do not properly conserve mass.  At root, the difficulty emerges because the excursion set formalism considers the trajectories of independent points in the IGM rather than grouping them into discrete halos which can then merge in a well-defined manner.  The mathematically self-consistent solution is to invert the Smoluchowski equation, but that is numerically challenging.  Fortunately, despite their inconsistencies extended Press-Schechter halo merger rates do offer useful insights into galaxy formation.  We hope that the bubble merger rate given by equation (\ref{eq:mergerate}) does the same.  To correct for the asymmetric merger kernel, we will use the symmetrized version $2 Q_{\rm sym}(m_1,m_2) \equiv Q(m_1,m_2) + Q(m_2,m_1)$.  Note that the aforementioned difficulties may be less severe for ionized bubbles because of their small-mass cutoff.

\begin{figure}
\begin{center}
\resizebox{8cm}{!}{\includegraphics{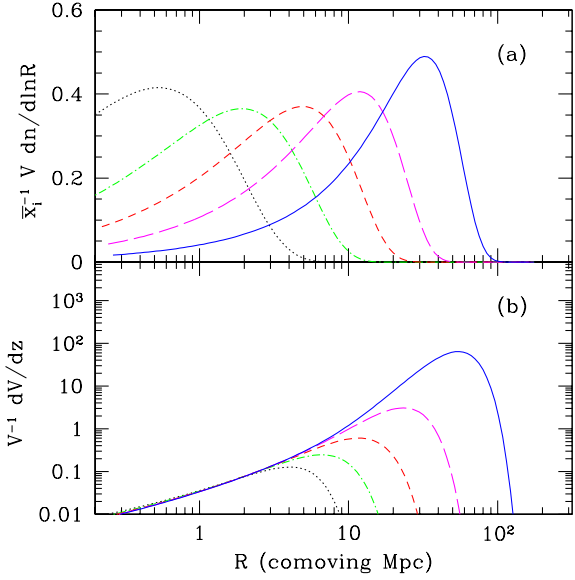}}\\%
\end{center}
\caption{\emph{(a)}: Size distribution of bubbles at $z=12$ if $\zeta=10,\, 20,\, 30,\, 40,$ and $50$ (dotted, dot-dashed, short-dashed, long-dashed,and solid curves, respectively).  \emph{(b)}: Merger rates at $z=12$ for $m_1=10^{14} \Msun$. The different curves correspond to the same choices for $\zeta$ as in the top panel.  The ordinate is defined in eq. (\ref{eq:volacc}).}
\label{fig:merger}
\end{figure}

Figure~\ref{fig:merger} shows some sample merger rates at $z=12$.  The top panel shows the bubble size distribution if $\zeta=10,\, 20,\, 30,\, 40,$ and $50$ (from left to right).  These correspond to $\bxio=0.17,\, 0.34,\, 0.51,\, 0.68$, and $0.85$.  The bottom panel shows the merger rate for bubbles with $m_1=10^{14} \Msun$.  The ordinate is
\bqa
V(m_1) ^{-1} \frac{\deriv V}{\deriv z} & \equiv & \frac{V(m_2)}{V(m_1)} \ m_2 n_b(m_2,z) \nonumber \\ & & \times \, Q_{\rm sym}(m_1,m_2,t) \ \left| \frac{\deriv t}{\deriv z} \right|,
\label{eq:volacc}
\eqa
which gives the fractional volume accretion rate by which bubbles of mass $m_1$ grow through mergers with bubbles of mass $m_2$ (in logarithmic mass intervals).  Clearly, regardless of $\bxio$, ionized regions tend to merge most rapidly with bubbles slightly above the current characteristic radius, so mergers with larger systems dominate the accreted mass regardless of the bubble of interest.  The total mass accretion rate increases rapidly throughout reionization because there is more gas in ionized bubbles at late times and because each bubble is larger.  Thus \hii regions grow at an accelerating rate as $\bxio \rightarrow 1$ (see also Fig.~\ref{fig:pmfp}\emph{b}).  Finally, note that the fractional accretion rate approaches a fixed asymptotic form for small bubbles.  This is because equation (\ref{eq:mergerate}) becomes independent of $m_2$ ($m_T \approx m_1$ and $\sigma_T^2 \approx \sigma_1^2$) when $m_2 \ll m_1$.

\begin{figure}
\begin{center}
\resizebox{8cm}{!}{\includegraphics{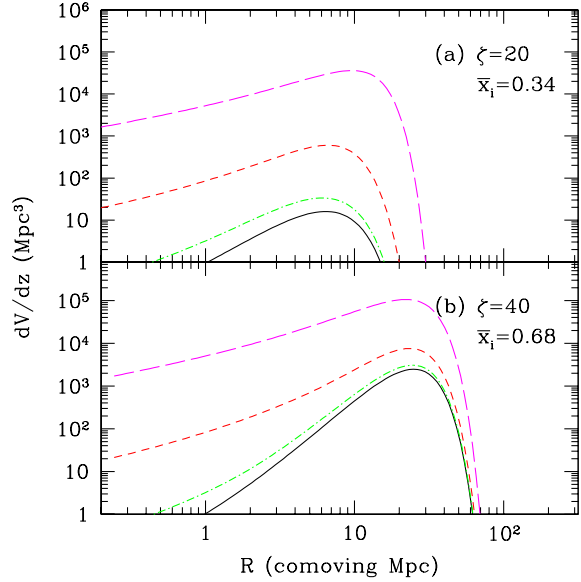}}\\%
\end{center}
\caption{Merger rates at $z=12$ for $m_1=10^{11},\, 10^{12},\, 10^{14}$, and $10^{16} \Msun$ (solid, dot-dashed, short-dashed, and long-dashed curves, respectively).  The ordinate is defined in eq. (\ref{eq:volacc}).  \emph{(a)}: $\zeta=20$.  \emph{(b)}: $\zeta=40$.  }
\label{fig:merger_mass}
\end{figure}

Figure~\ref{fig:merger_mass} shows similar results, except we fix $\zeta$ and vary $m_1$.  We show the volume accretion rate (in dimensional comoving units), again as defined in equation (\ref{eq:volacc}).  The solid, dot-dashed, short-dashed, and long-dashed curves take $m_1=10^{11},\, 10^{12},\, 10^{14}$, and $10^{16} \Msun$, respectively.  These have $R_1=0.84,\,1.8,\,8.4,$ and $38.9$ comoving Mpc.  The two panels assume $\zeta=20$ ($\bxio=0.34$) and $\zeta=40$ ($\bxio=0.68$); Figure \ref{fig:merger}\emph{a} shows the corresponding $n_b(m)$.  Interestingly, large bubbles grow faster, in absolute terms, than small bubbles.  The disparity is greatest for the accretion of small bubbles by large ionized regions:  apparently, as these expand they swallow up small bubbles quite rapidly.  On the other hand, the rate at which a given bubble merges with large \hii regions has a much weaker mass dependence.  In particular, if we had normalized these curves by $V(m_1)^{-1}$ (as in Fig.~\ref{fig:merger}), we would find that small bubbles grow fractionally at much faster rates than large bubbles, because both tend to merge with ionized regions of size $R_{\rm char}$.  

These merger rates have interesting implications for our understanding of the overlap process.  Because $R_{\rm char}$ is well-defined, mergers with objects of this size dominate bubble growth -- regardless of the bubble under consideration.  This implies that small bubbles, which grow in voids, do not merge with each other:  instead they overlap only when a large bubble, from a distant cluster of sources, sweeps past them.  Such behavior is consistent with the ``inside-out" nature of reionization.  It also helps to explain the rapidly increasing $R_{\rm char}$ seen in Figure~\ref{fig:pmfp}\emph{b}:  most of the universe lies inside bubbles near $R_{\rm char}$, which merge with bubbles of slightly greater sizes at ever-increasing rates.  

\begin{figure}
\begin{center}
\resizebox{8cm}{!}{\includegraphics{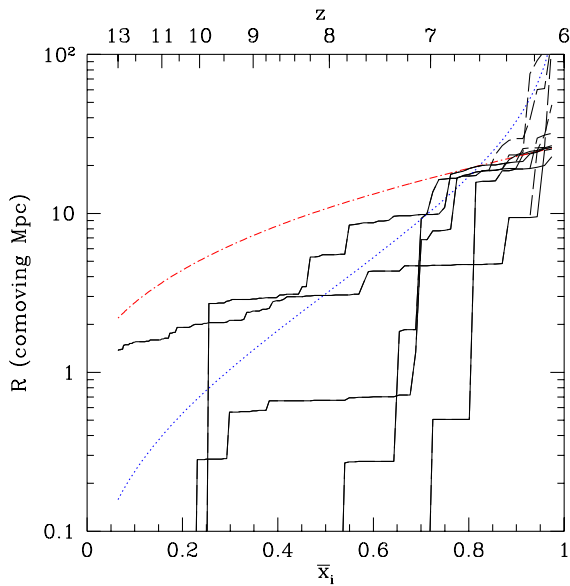}}\\%
\end{center}
\caption{Bubble histories for several randomly generated trajectories.  The abscissa shows the bubble radius surrounding each point as a function of $\bxio$ and $z$; we fix $\zeta$ so that reionization completes at $z=6$.  The solid lines include the recombination limit (see \S \ref{recomb} below) while the dashed curves use the unmodified FZH04 model (the two are coincident until $\bxio \ga 0.9$).  The dotted and dot-dashed lines show $R_{\rm char}$ and $R_{\rm max}$, respectively.}
\label{fig:rw}
\end{figure}

Most importantly, this picture also affects our interpretation of the evolving radiation background.  Figure~\ref{fig:rw} illustrates how.  We have generated an ensemble of random walks in the $(\sigma^2,\delta)$ plane and tracked their bubble histories.  We fixed $\bxio=1$ at $z=6$, assumed constant $\zeta$, and compared each random walk to our FZH04 barrier at intervals of $\Delta z=0.05$. We begin at $z=13$, when $\bxio = 0.065$.  The solid and dashed curves are for models with and without recombinations, respectively (to be discussed in \S \ref{recomb} below); for now the middle stages are most relevant.  The dotted line shows $R_{\rm char}$ for reference.  Even though $\bxio$ grows smoothly in this model, the process is extremely discontinuous if viewed from any given point:  although the local emissivity grows only slowly, the point joins progressively larger bubbles as time goes on.  Moreover, large bubbles always dominate the volume accretion rate, so each merger is ``major" (i.e., it typically increases the volume by a factor of two or much more) and causes a substantial jump in the bubble radius.  Note that this is even true when bubbles contain many ionizing sources -- the variance does \emph{not} decrease like a Poisson process.  Because (neglecting recombinations) the bubble radius is also the horizon out to which galaxies are visible, this in turn implies that the local ionizing background evolves through a series of sharp, large amplitude jumps.   Obviously we must be careful not to interpret these \emph{local} jumps in the ionizing background as signatures of \emph{global} overlap.  This contrasts with the standard picture from simulations \citep{gnedin00,razoumov02}, in which a jump in the ionizing background was attributed to the completion of reionization.  These simulations were limited to small boxes with sizes $\sim 1$--$10 \Mpc$.  As such, they are much smaller than the typical \hii regions at the end of reionization, and the simulation jumps occurred when a bubble encompassed the entire box.  But because the small boxes are not characteristic of the entire universe (\citealt{barkana04-fluc}; FZH04), this event need not correspond to global overlap.  Our models show that interpreting such jumps is more subtle.

\section{The IGM Density Distribution}
\label{density}

\subsection{The MHR00 model}
\label{mhr}

To this point, we have essentially ignored recombinations ($\zeta$ includes only a factor accounting for the mean integrated number per baryon).  In the following sections, we will consider how recombinations in an inhomogeneous IGM affect the bubble topology.  To begin, we will review a simple analytic model for the IGM density distribution and its ionization state originally due to MHR00.  They used numerical simulations to measure the volume-weighted density distribution of IGM gas, $P_V(\Delta)$, where $\Delta=\rho/\bar{\rho}$, and found a good fit at $z \sim 2$--$4$ with 
\bq
P_V(\Delta) \, \deriv \Delta = A_0 \Delta^{-\beta} \exp \left[ - \frac{(\Delta^{-2/3} - C_0)^2}{2(2 \delta_0/3)^2} \right] \, \deriv \Delta.
\label{eq:pvd}
\eq
Intuitively, the underlying Gaussian density fluctuations are modified through nonlinear void growth and a power law tail at large $\Delta$.  MHR00 argued that the form could be extrapolated to higher redshifts in the following way.  First, $\delta_0$ essentially represents the variance of density fluctuations smoothed on the Jeans scale for an ionized medium; thus $\delta_0 \propto (1+z)^{-1}$ at  high redshifts, as their simulation found for $z=2$--$4$.  The power-law exponent $\beta$ determines the behavior at large densities; for isothermal spheres, it is $\beta=2.5$.  The remaining constants ($A_0$ and $C_0$) can be set by demanding proper mass and volume normalization.

The next step is to determine how this gas is ionized.  MHR00 assumed that there exists a density threshold $\Delta_i$ such that all the gas with $\Delta<\Delta_i$ is ionized while all gas with $\Delta > \Delta_i$ remains neutral.  This provides a reasonable description of shielding in dense regions, if those regions can be considered to be isolated clumps in which the density increases inwards.  In that case, the radiation field will ionize the outskirts of the cloud until $\tau \approx 1$.  Because of the density gradient, this skin corresponds to our threshold $\Delta_i$.  

Given this simple prescription, the recombination rate of ionized gas, denoted $A$ (note the change from MHR00 in order to avoid confusion with the bubble radius), is
\bq
A(\Delta_i) = A_u  \int_0^{\Delta_i} \deriv \Delta \, P_V(\Delta) \, \Delta^2 \equiv A_u C,
\label{eq:arate}
\eq
where $A_u$ is the recombination rate per hydrogen atom in gas at the mean density,
\bq
A_u = \alpha_A(T) \bar{n}_e,
\label{eq:au}
\eq
and $C$ is the effective clumping factor.  Here, following MHR00, we use the case-A recombination rate (see also \citealt{miralda03}), $\alpha_A=4 \times 10^{-13} \recunits$ (at $T=10^4 \kel$).  Case-A is appropriate because most recombinations occur in gas with $\Delta \sim \Delta_i$, which presumably lies on the edge of dense neutral clouds.  Thus the secondary ionizing photons that follow recombinations to the ground state will find themselves in the dense, neutral clouds and do not ionize the diffuse IGM in which we are interested.  In this picture $C$ increases rapidly as the universe becomes more and more ionized; thus, the first phase of reionization is more efficient than later stages, when we must clear out dense, rapidly-recombining blobs of gas.  

MHR00 also offer a prescription for determining $\lambda_i$, the mean free path of ionizing photons.  In their model, it equals the mean distance between clumps with $\Delta>\Delta_i$ along a random line of sight, which is approximately
\bq
\lambda_i = \lambda_0 [1 - F_V(\Delta_i)]^{-2/3}.
\label{eq:mfp-mhr}
\eq
Here $F_V(\Delta_i)$ is the fraction of volume with $\Delta < \Delta_i$ and $\lambda_0$ is a (redshift-dependent) normalization factor.  Formally, this expression is valid only if the number density and shape (though not total cross section) of absorbers is independent of $\Delta_i$.  This is obviously not true in detail for the cosmic web.  However, MHR00 found that it provided a good fit to numerical simulations at $z=2$--$4$ if we set $\lambda_0 H(z) = 60 \kms$ (in physical units).  We will extrapolate the same prescription to higher redshifts.  

While it does a remarkable job matching many aspects of the Ly$\alpha$ forest at moderate redshifts, applying the MHR00 model to high redshifts involves a number of potentially dangerous simplifications and extrapolations.  In the Appendix, we examine the model in some more detail and illuminate many of its features.  For example, we show that equation (\ref{eq:mfp-mhr}) probably overestimates $\lambda_i$ by up to a factor $\sim 2$ because of accumulated photoelectric absorption by low-column density systems.  We urge the reader to keep in mind the approximate nature of our model.  We emphasize that better measurements with high-resolution simulations will be useful in the future.

\subsection{A minihalo model}
\label{minihalo}

One possibly unwarranted assumption of MHR00 is the explicit smoothing of the IGM on the Jeans scale for $T=10^4 \kel$ gas.  While appropriate well after reionization, this may not apply during that epoch.  If the IGM remains cool before reionization, small collapsed objects called ``minihalos" could form; they will evaporate during reionization \citep{barkana99,shapiro04} but may constitute a substantial photon sink before they are completely destroyed \citep{haiman01-mh,barkana02-mh,iliev05}.  However, the importance of minihalos remains controversial:  their formation relies on fluctuations on extremely small scales, and even a modest amount of heating will strongly suppress their abundance \citep{oh03-entropy}.   For transparency, and because minihalo formation is so unconstrained, we therefore use an extremely simple model for a universe in which minihalos control the recombination rate.  We suppose that a fraction $f_{\rm mh}$ of the IGM has collapsed into dense blobs with mass $M_{\rm mh}$ and mean density $\Delta_{\rm mh} \approx 18 \pi^2$.  We will assume that these minihalos remain neutral, randomly distributed sinks for ionizing photons.  Then the number density of minihalos is $n_{\rm mh}=f_{\rm mh} \bar{\rho}/M_{\rm mh}$, each with radius $R_{\rm mh} \propto (M_{\rm mh}/\Delta_{\rm mh})^{1/3}$.  The mean free path of an ionizing photon is
\bqa
l_{\rm mh} & \equiv & \frac{1}{\pi n_{\rm mh} R_{\rm mh}^2} \approx 15.7 \left( \frac{M_{\rm mh}}{10^6 \Msun} \right)^{1/3} \ \left( \frac{0.05}{f_{\rm mh}} \right) \nonumber \\
& & \times  \left( \frac{\Delta_{\rm mh}}{18 \pi^2} \right)^{2/3} \ \left( \frac{\Omega_m h^2}{0.15} \right)^{-1/3} \Mpc
\label{eq:lmh}
\eqa
in comoving units.  The fraction of ionizing photons surviving to a distance $R$ from a source is $f_s = 1 -\exp (-R/l_{\rm mh})$.

\section{Bubbles with Recombinations}
\label{recomb}

\subsection{The recombination barrier}
\label{barrier}

We are now in a position to incorporate recombinations into the FZH04 model.  Our fundamental underlying assumption is that the bubble distribution depends on the underlying density field:  \hii regions correspond to (large-scale) overdense regions of the universe.  We will therefore need to write the recombination rate as a function of the (smoothed) mean density.   Conceptually, we must modify equation (\ref{eq:arate}) in two ways.  First, the recombination rate per baryon increases with the mean density of the region:  $A_u \propto (1+\delta)$.  Second, denser regions can be thought of as sub-universes with a larger matter density.  In such regions, structure formation proceeds more rapidly than average [$H(z) \propto \sqrt{\Omega_m}$ at high redshifts], so they are slightly clumpier than average.  This adds a second layer to the clumpiness calculation:  in principle $P_V(\Delta) \rightarrow P_V(\Delta|\delta,R)$.  For simplicity, and because the overdensities of bubbles are modest during most of reionization (one implication of their large sizes), we will neglect the latter effect and assume $P_V(\Delta)$ is independent of environment.  

Now suppose that we have a bubble of size $R$.  In order for the bubble to grow, ionizing photons must be able to reach its edge.  This demands that $\lambda_i \geq R$, setting the threshold $\Delta_i$ to which we must have ionized the gas inside the bubble through equation (\ref{eq:mfp-mhr}).  Obviously, larger bubbles contain more highly ionized material, because their photons must propagate farther.  At the same time, the ionized outskirts of dense clumps are recombining, and we must counteract these recombinations in order for any photons to reach the edge of the \hii region.  Thus for bubbles to continue growing, the rate at which they ionize their constituent gas must exceed the rate at which it recombines:
\bq
\zeta \frac{\deriv f_{\rm coll}(\delta,R)}{\deriv t} > A_u C'(\delta,R).
\label{eq:recombbarrier}
\eq
The condition $\lambda_i = R$ sets $\Delta_i$; with our assumption that $P_V(\Delta)$ is independent of the mean bubble density, we can then write
\bq
C'(\delta,R) = C(R)(1+\delta).
\label{eq:C}
\eq
Note that bubbles are not completely ionized in this picture, because each contains some fraction of dense neutral gas.  The quantity $\bxio$ should be thought of as the fraction of matter contained in \hii regions rather than the fraction of hydrogen atoms that are actually ionized.

Equation (\ref{eq:recombbarrier}) differs from the FZH04 condition $f_{\rm coll} > \zeta^{-1}$, which requires only that the sources inside produce enough photons to ionize every hydrogen atom when integrated over the history of the universe (or at best to counteract the cumulative number of recombinations).  The condition of equation (\ref{eq:recombbarrier}) instead requires that the sources produce enough ionizing photons to counteract recombinations \emph{at that instant}.  Clearly these are two limiting cases:  in reality the sources must produce enough photons to ionize each atom \emph{plus} cancel ongoing recombinations.  
It is thus nothing more than the canonical Str{\" o}mgren radius, in which recombinations exactly balance ionizations.
Separating the problem into these two limits is useful so long as one or the other clearly dominates over most of parameter space, which as we shall see below is indeed the case.  We note that MHR00 made the same approximation in their Figure 7; our treatment simply extends theirs to include the inhomogeneous distribution of (clustered) sources throughout the universe.

\begin{figure}
\begin{center}
\resizebox{8cm}{!}{\includegraphics{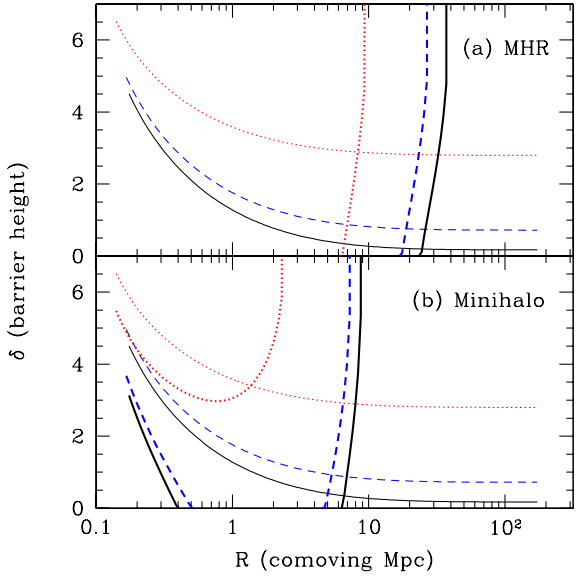}}\\%
\end{center}
\caption{Barrier height as a function of comoving radius.  The solid, dotted, and dashed curves have $\bar{x}_i=0.95,\,0.82$, and $0.49$, respectively, at $z=6$.  Within each set, the thin curves show $\delta_x$ (from FZH04) and the thick curves show $\delta_r$ (from this paper).  \emph{(a)}:  Uses MHR00 to compute $\delta_r$.  \emph{(b)}:  Uses the minihalo model described in the text to compute $\delta_r$.  }
\label{fig:dxz6}
\end{figure}

Equation (\ref{eq:recombbarrier}) can be solved numerically to yield the minimum $\delta_r$ as a function of scale $M$ required for ionizing sources to counteract recombinations inside a bubble.  In the excursion set formalism, a trajectory must then cross $\delta_b  \equiv {\rm max}(\delta_x,\delta_r)$ in order to be part of an ionized bubble.  We compare these barriers in Figure~\ref{fig:dxz6}\emph{a}.   The three monotonically decreasing curves show $\delta_x$, the FZH04 barrier, at $\bxio=0.95,\,0.82$, and $0.49$ (solid, dotted, and dashed lines, respectively).\footnote{As shown in FZH04, these barriers are nearly linear when expressed as functions of $\sigma^2(m)$.} The thick nearly vertical curves show $\delta_r$ for the same neutral fractions.  On small scales, recombinations provide a much less stringent limit than simply ionizing the gas:  small bubbles have a low density threshold $\Delta_i$, because the photons need not propagate far.  Thus the recombination time inside each bubble is long.  However, as we increase the required $\Delta_i$, $A$ increases more rapidly than the collapse fraction (see, e.g., Fig. 2 of MHR00) and $\delta_r$ becomes the limiting factor.  The steepness of $\delta_r$ is a result of the power-law form of equation (\ref{eq:pvd}) at large $\Delta$, which assumes that dense clumps are isothermal spheres, $\rho \propto r^{-2}$.  Increasing $\lambda_i$ requires a substantial increase in $\Delta_i$ and hence in the recombination rate.  Meanwhile, increasing $\deriv f_{\rm coll}/\deriv z$ is difficult, so the required underlying density grows rapidly.  The crossover between the two limits depends on the ionized fraction, but it typically occurs at $R \ga 10 \Mpc$.  Figure~\ref{fig:pmfp} shows that the bubbles do not approach this scale until $\bxio \ga 0.75$, so the recombination limit only affects the bubble size distribution late in reionization.  
: explanation of the title
We thus see that recombinations play an increasingly crucial role as the bubbles grow larger and larger.  They constitute a progressive tax on the ionizing sources, with richer clusters paying more because their bubbles are intrinsically larger.

Figure~\ref{fig:dxz6} suggests that the crossover scale $R_{\rm max}$, where $\delta_x=\delta_b$, does evolve as reionization proceeds.  However, the rate of evolution is quite modest.  We show this explicitly in Figure~\ref{fig:pmfp}\emph{b}.  The thin curves (shown for $z=6$ and $9$) plot the maximum of $R_{\rm char}$ and $R_{\rm max}$; the turnover at $\bxio \ga 0.9$ is due to the recombination limit.  Thus, \hii regions continue to grow after recombinations become important, but only slowly -- the evolution toward $\bxio \approx 1$ is completely different from a case with no recombinations, in which the bubble growth is unbounded.  This slow growth is similar to what we expect in the post-overlap universe, because bubble mergers no longer increase the ionizing background (see \S \ref{overlap}).

We can apply similar arguments to a universe in which minihalos control the mean free path.  In this case, the fraction of photons propagating to the edge of the \hii region is $f_s$ (recall we assume that any photon impacting a minihalo is lost irrevocably).  Thus, the efffective clumpiness is
\bq
C_{\rm mh}(R) = (1-f_{\rm mh})^2 \exp(R/l_{\rm mh}),
\label{eq:clump-mh}
\eq
where the first factor describes the fraction of mass remaining in the diffuse IGM.  We again take $C_{\rm mh}'(\delta,R)=(1+\delta)C_{\rm mh}(R)$, which assumes that minihalo formation is unbiased. That is certainly not true, but given the simplicity of our treatment it suffices.  (Note also that \citealt{iliev05} argue that the effective bias of minihalos around a source is more or less independent of scale, which implies that we are not ignoring any crucial scale dependence.)  Figure~\ref{fig:dxz6}\emph{b} shows $\delta_r$ for our minihalo model with $f_{\rm mh}=0.05$ and $M_{\rm mh}=10^6 \Msun$.  We see that these parameters provide a more stringent constraint on the bubble sizes than MHR00:  the maximum size is limited to $\sim 1 \Mpc$ during the middle stages of reionization and somewhat under $\sim 10 \Mpc$ toward the end.  Thus, in principle, minihalos can strongly affect the topology of reionization; however, at least in this simple treatment, they do not completely destroy the large-scale fluctuations.  Clearly,  measurements of the distribution of ionized gas can provide powerful limits on their formation.

\begin{figure}
\begin{center}
\resizebox{8cm}{!}{\includegraphics{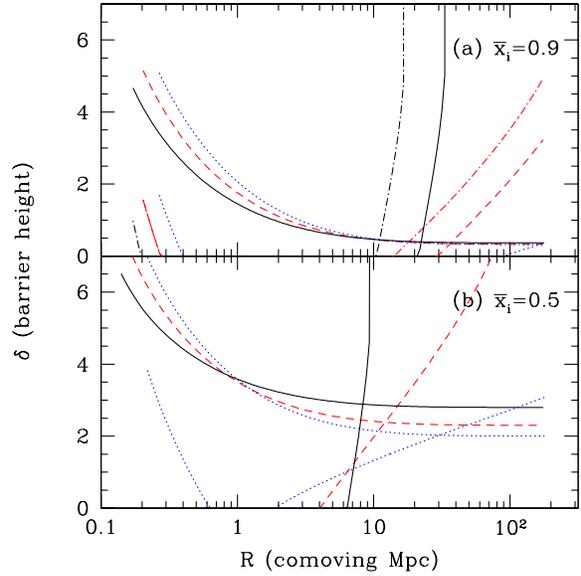}}\\%
\end{center}
\caption{Barrier height as a function of comoving radius.  The solid, dashed, and dotted curves in each panel are for $z=6,\,9$, and $12$;  the monotonically decreasing curves show $\delta_x$, while the others are the recombination barriers $\delta_r$ (assuming the MHR00 model).  \emph{(a)}:  $\bxio=0.9$.  The dot-dashed curves set $\lambda_0 H(z)=30 \kms$ for $\delta_r$ (see text).  \emph{(b)}:  $\bxio=0.5$. }
\label{fig:dxz}
\end{figure}

Figure~\ref{fig:dxz} shows the barriers for $z=6,\,9$ and $12$ (solid, dashed, and dotted curves, respectively) at fixed $\bxio=0.5$ and $0.9$.  As pointed out by FZH04, the simple $\delta_x$ barrier changes relatively little with redshift so long as we fix $\bxio$; the only difference is the slow decline of $m_{\rm min}$ with cosmic time.  However, recombinations do evolve strongly with redshift, because they depend directly on the clumpiness.  Interestingly, the MHR00 $\delta_r$ not only crosses $\delta_x$ at different scales as a function of $z$, but it also changes slope.  Although the parametric form of $P_V(\Delta)$ remains constant with redshift, $\delta_r$ is sensitive to a different portion of $P_V(\Delta)$ at each redshift.  Because the clumpiness decreases at high redshifts, the decline in the net recombination rate actually allows the bubbles to grow larger, even though $A_u$ itself increases.  In any case, we note that if the MHR00 model is accurate, recombinations will only limit the bubbles on tens of Mpc (or even 100 Mpc) scales:  with this model, reionization is inevitably accompanied by large-scale fluctuations in the ionizing background.

As described in the Appendix, the MHR00 prescription may overestimate the mean free path of ionizing photons by up to a factor of two.  The dot-dashed curves in Figure~\ref{fig:dxz}\emph{a} show that setting $\lambda_0 H(z) = 30 \kms$ would, not surprisingly, decrease the maximum bubble size by about the same factor (they are otherwise constructed in the same way as the usual $\delta_r$ curves).  Because of the uncertainties inherent to existing models of the IGM density field we cannot quantitatively determine the maximum bubble size with much confidence; however, eliminating fluctuations on scales $\ga 10 \Mpc$ requires doing a fair amount of violence to the model, such as adding substantial numbers of minihalos.

The scale at which $\delta_r=\delta_x$ is also quite sensitive to the other input parameters. This is a consequence of its steepness:  order unity changes to the right-hand side of equation (\ref{eq:recombbarrier}) require corresponding changes to $\deriv f_{\rm coll}/\deriv z$, which (due to its weak scale dependence) cannot be done without significantly altering the bubble size. For example, in equation (\ref{eq:au}), we have assumed that case-A recombination is appropriate, because the re-emitted photons will generally be lost inside dense neutral regions.  If we instead assume case-B recombination, $A_u(B) \approx 0.6 A_u(A)$.  We find that this imposes a limiting size about twice as large as the case-A values at $z=6$.  This is another reason why numerical simulations will eventually be required to fully understand reionization.

\subsection{The bubble size distribution with recombinations}
\label{bubsize}

We have argued that treating the excursion set barrier as the maximum of $\delta_x$ and $\delta_r$ is a good approximation for the effects of recombinations on bubble growth.  Armed with this barrier, we can now consider its effects on the bubble size distribution.  From Figures~\ref{fig:dxz6} and \ref{fig:dxz}, it is obvious that recombinations are unimportant until some scale $R_{\rm max}$ is reached, where $\delta_x=\delta_r$ and above which $\delta_r$ rises rapidly.  The slope in this regime depends on redshift, but it is always much steeper than the nearly flat $\delta_x$ barrier.  Thus, the primary effect of recombinations will be to prevent bubbles growing larger than $R_{\rm max}$.  Early in reionization, when $R_{\rm max} \gg R_{\rm char}$, recombinations can be safely ignored.  Later on, we expect the bubbles to saturate near $R_{\rm max}$.  The trajectories in Figure~\ref{fig:rw} bear out this intuition.  The dot-dashed curve shows $R_{\rm max}(z)$.  The solid and dashed lines show the bubble histories with and without recombinations, respectively.  Early on, the two sets of curves are coincident because recombinations have no effect on the bubbles.  However, when $\bxio \ga 0.9$, $R_{\rm char} > R_{\rm max}$ and the trajectories depart from one another.  Without recombinations, the bubbles continue to grow; with them, they saturate and cluster around $R_{\rm max}$.

As in FZH04, we must approximate the barrier with a simple form in order to estimate $n_{\rm rec}(m)$, the bubble size distribution with recombinations.  The most straightforward choice is to assume that $\delta_r$ is simply a vertical line that crosses $\delta_x$ at $R_{\rm max}$.  This is actually an excellent approximation in the $(\sigma^2,\delta)$ plane, because $\sigma$ is already so small at the relevant scales.  We can then solve for $n_{\rm rec}(m)$ by noting that there is no barrier at all for $R>R_{\rm max}$.  The probability distribution at this scale will then be a simple gaussian:
\bq
p(\delta | R_{\rm max}) \, \deriv \delta = \frac{1}{\sqrt{2 \pi} \sigma_{\rm max}} \ \exp \left( \frac{-\delta^2}{2 \sigma_{\rm max}^2} \right) \, \deriv \delta,
\label{eq:pdel}
\eq  
where $\sigma_{\rm max} \equiv \sigma(R_{\rm max})$.  Our barrier is a vertical line beginning at $B(R_{\rm max})$, so any trajectory that lies above the FZH04 barrier (when evaluated on the saturation scale) will be incorporated into a bubble with $R=R_{\rm max}$.  The number density of such bubbles is then
\bq
N_{\rm rec}(m_{\rm max}) =  \frac{\bar{\rho}}{2 m_{\rm max}} \ {\rm erfc} \left[ \frac{B(R_{\rm max})}{\sqrt{2} \sigma_{\rm max}} \right].
\label{eq:fcl}
\eq
(note that this is a true number density with units inverse comoving volume).

Trajectories with $\delta(R_{\rm max})<B(R_{\rm max})$ continue their random walks until they cross the FZH04 barrier on smaller scales.  The result must be independent of the trajectory at $R>R_{\rm max}$; we only care about its value where the barrier begins.  The resulting mass function is [excluding the true recombination-limited bubbles of eq. (\ref{eq:fcl})]:
\bq
n_{\rm rec}(m,z) =  \int_{-\infty}^{B(R_{\rm max})} \deriv \delta \ p(\delta | R_{\rm max}) \ n_b(m,z|\delta,m_{\rm max},z),
\label{eq:nmrec}
\eq
where $n_b(m,z|\delta,m_{\rm max},z)$ is the conditional mass function for a trajectory that begins its random walk at the point $(\sigma_{\rm max}^2,\delta)$.  This function is identical to equation (\ref{eq:prog}) except with the replacement $B(M_b,z_b) \rightarrow \delta$ because the trajectory starts at an arbitrary density rather than in a specified bubble.  In other words, the net mass function is the weighted average of the conditional mass functions evaluated over all densities smoothed on the scale $R_{\rm max}$.

\begin{figure}
\begin{center}
\resizebox{8cm}{!}{\includegraphics{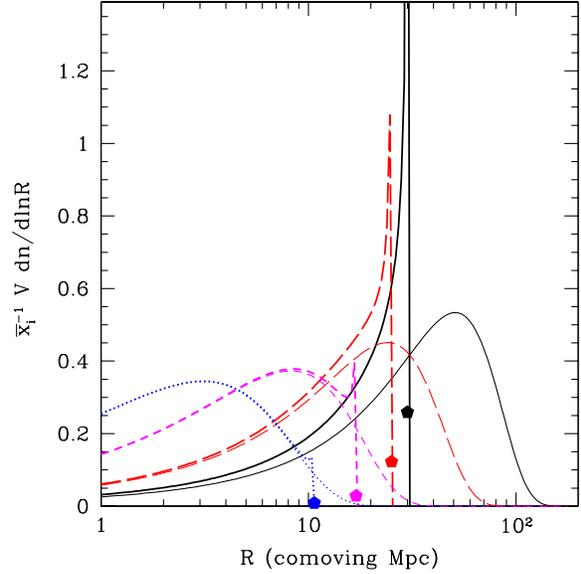}}\\%
\end{center}
\caption{Bubble size distributions at $z=8$ for the unmodified FZH04 model (thin curves) and including recombinations (thick peaked curves).  The dotted, short-dashed, long-dashed, and solid curves have $\bxio=0.51,\,0.68,\,0.84$, and $0.92$, respectively.  The filled hexagons show the fraction of the ionized volume in bubbles with $R=R_{\rm max}$ for the model with recombinations.}
\label{fig:recbub}
\end{figure}

Although the integral in equation (\ref{eq:nmrec}) can be solved analytically, the result is complicated and far from illuminating.  We therefore simply show some example size distributions in Figure~\ref{fig:recbub}.  The thick and thin curves show the sizes with and without recombinations, respectively.  We take $z=8$ and show results for $\bxio=0.51,\,0.68,\,0.84$, and $0.92$ (dotted, short-dashed, long-dashed, and solid curves, respectively).  If the characteristic size is much smaller than the recombination limit, $n_{\rm rec} \approx n_b$ except for the large bubble tail (e.g., the dotted curve).  As $R_{\rm char} \rightarrow R_{\rm max}$, the bubbles pile up near that scale.  Without recombinations, trajectories in slightly overdense regions tend to cross $\delta_x$ on scales slightly larger than $R_{\rm char}$.  With recombinations added, that is no longer possible, delaying their first crossing until just after $R_{\rm max}$.  When $R_{\rm char} \gg R_{\rm max}$, the pileup becomes more significant and a large fraction of the universe lies in bubbles at or just below the recombination limit.

The filled hexagons in Figure~\ref{fig:recbub} show the fraction of the ionized volume in bubbles with $R=R_{\rm max}$.  This is \emph{not} equal to the fraction of the volume in bubbles with $R>R_{\rm max}$ for the pure FZH04 model, because some trajectories cross $\delta_x$ at large scales but then fall below it (i.e., small voids in wide overdensities).  If the FZH04 barrier were constant, exactly half as many trajectories would lie in saturated bubbles as lie in large bubbles for the pure FZH04 model.\footnote{This is the same factor of two that bedeviled the original \citet{press74} derivation of the halo mass function.  It can be most easily understood through the excursion set formalism, because the barrier absorbs both those trajectories that continue above it and those that reflect off of it \citep{chandrasekhar43,bond91}.}  With our linear barrier, the difference is slightly larger [compare our equation (\ref{eq:fcl}) with equation (11) of \citealt{mcquinn05}].  The other half of the trajectories that would be in large bubbles continue to smaller scales; most fill up the peak just below $R_{\rm max}$.

For the same reason, the total ionized fraction with recombinations is actually slightly smaller than $\zeta f_{\rm coll,g}$:  a small fraction of trajectories, corresponding to deep voids embedded in large scale overdensities, would have crossed the FZH04 barrier \emph{only} on scales $R>R_{\rm max}$.  Fortunately, this is a small correction -- most likely less important than the residual neutral gas within the bubbles -- so we will not examine it further here.

\subsection{The ``time of overlap"}
\label{overlap}

Thus, recombinations impose a saturation radius $R_{\rm max}$ beyond which bubbles cannot grow, as ionizing photons are consumed internally rather than reaching the edge of the bubble.  From the point of view of a single point in the IGM, reionization is complete once that point joins such a Str{\" o}mgren sphere.  Even if another \hii region forms contiguous to its host bubble, the ionizing background will change only slightly.  ``Overlap" effectively occurs at different times throughout the IGM and should be considered a \emph{local} process.  
After saturation, the mean free path (and hence the ionizing background) grows much more slowly (see Figs.~\ref{fig:pmfp}\emph{b} and \ref{fig:rw}), \emph{even though the universe as a whole is not fully ionized}.  According to the picture of \citet{gnedin00}, such slow growth is characteristic of the post-overlap phase.  Instead, we argue that the it is characteristic of saturation, which occurs throughout the end of reionization.  Our model provides a well-defined way to compute this distribution of overlap times.  We let $x_{\rm rec}$ be the fraction of points contained in bubbles with $R>R_{\rm max}$ (computed using the standard FZH04 barrier).  As shown in the previous section, this is essentially the same as the fraction of trajectories in recombination-limited (or nearly recombination-limited) bubbles.  

\begin{figure}
\begin{center}
\resizebox{8cm}{!}{\includegraphics{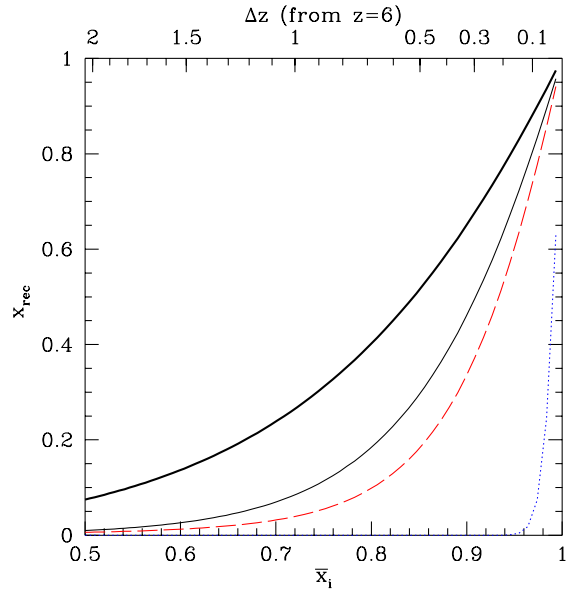}}\\%
\end{center}
\caption{Fraction of the universe in Str{\" o}mgren spheres; i.e., contained in FZH04 bubbles larger than $R_{\rm max}$.  Solid, dashed, and dotted curves are for $z=6,\,9$, and $12$, respectively.  The thin lines use the MHR00 density model. The thick solid curve assumes that $\lambda_i$ is half the MHR00 value at $z=6$.  The top axis shows the redshift offset from the time of full ionization (if that occurs at $z=6$), assuming that $\zeta$ is fixed.}
\label{fig:qrec}
\end{figure}

Figure~\ref{fig:qrec} shows $x_{\rm rec}$ as a function of $\bxio$ at three different redshifts ($z=6,\,9$, and $12$ for the thin solid, dashed, and dotted curves, respectively) assuming the MHR00 model for the mean free path.  In all cases, only a small fraction of the IGM reaches saturation early in reionization:  none of these large bubbles form before $\bxio=0.5$, and at least 50$\%$ of the universe reaches local overlap in the range $0.9<\bxio<1$.  This fraction increases at higher redshifts:  as seen in Figure~\ref{fig:dxz}, the lack of substructure in the high-redshift universe allows the bubbles to grow much larger at $z=12$ than at $z=6$, so the saturation limit occurs at much larger $R_{\rm max}$.  
This range in $\bxio$ corresponds to a small (but far from negligible) redshift interval.  To fix numbers, the top axis of Figure~\ref{fig:qrec} shows the redshift interval from complete reionization for the $z=6$ case, assuming a constant efficiency $\zeta$.  In this case, $x_{\rm rec} \approx 0.5$ when $\Delta z \approx 0.3$, with a long tail to even higher redshifts.  The redshift interval is somewhat smaller at higher redshifts, both because $R_{\rm max}$ is larger and because $f_{\rm coll}$ increases more rapidly (in relative terms).  
:  moved up from next paragraph
Interestingly, the interval increases if we decrease the Str{\" o}mgren radius:  the thick solid curve is the distribution in overlap times for $z=6$ in a model where we take the mean free path of ionizing photons to be one half of the MHR00 value of equation (\ref{eq:mfp-mhr}); according to the Appendix this is a reasonable guess for the cumulative effect of low-column density systems.  By decreasing $R_{\rm max}$, bubble saturation begins much earlier:  $x_{\rm rec} \approx 0.5$ at $\Delta z \approx 0.6$, with a larger tail.  Essentially, the bubbles saturates earlier in overdense regions, and underdense regions (which would otherwise be engulfed by bubbles propagating outward from filaments) take longer to catch up.

Interestingly, our model shows that we cannot identify the thickness of the reionization surface with the scale $R_{\rm max}$. Naively, one might expect the bubble size to provide a characteristic ``radius of curvature" for the reionization epoch, with larger bubbles resulting in more cosmic variance in the time of overlap.  \citet{wyithe04-var} argued that the width of the overlap surface comes from balancing causality with density fluctuations (quantified in a way similar to FZH04); essentially they equated $R_{\rm max}$ with the width of the overlap surface. Our model shows that this is an overestimate because of the spread in overlap times; in fact, a {\it smaller} $R_{\rm max}$ produces greater cosmic variance in the overlap era.  Thus, the difference in Lyman series transmission between the lines of sight to the two highest redshift quasars on $\sim 100$ Mpc lengthscales \citep{oh05,white05} does not necessarily require bubbles of size $\sim 100$ Mpc; instead, it implies that bubble growth saturated at considerably smaller length scales; according to our model, such variance could come from bubbles $\la 20 \Mpc$.  The strength of our model is that we can compute this variance quantitatively.

\subsection{The global recombination rate}
\label{recrate}

\begin{figure}
\begin{center}
\resizebox{8cm}{!}{\includegraphics{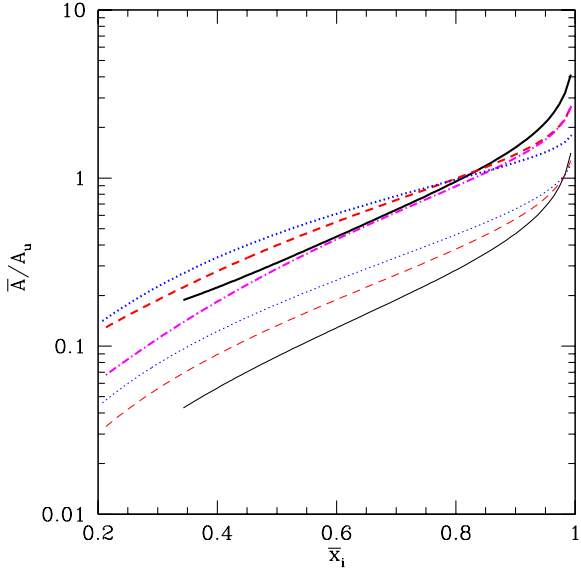}}\\%
\end{center}
\caption{Recombination rates for inhomogeneous reionization.  Each curve shows the net recombination rate $\bar{A}$ scaled to that for a homogeneous universe ($A_u$); the upper thick set includes the bubbles, while the lower thin set does not.  Solid, dashed, and dotted curves are for $z=6,\,9$, and $12$, respectively.  The dot-dashed curve includes the variable mean free path at $z=9$ but not the bubble overdensity.  Note that, in the lower redshift cases, curves begin at $\bxio=2 f_{\rm coll}$.}
\label{fig:abar}
\end{figure}

MHR00 pointed out that the global recombination rate will increase throughout reionization as the ionization fronts penetrate into denser and denser structures. With their assumption of a uniform ionizing background, they fixed $\Delta_i$ by requiring $F_V(\Delta_i)=\bxio$ (note that we treat $\bxio$ as a volume-weighted quantity here).  Then the global recombination rate $\bar{A}$ follows directly from equation (\ref{eq:arate}).  We show the resulting recombination rates, scaled to that of a homogeneous IGM ($A_u$), by the lower thin set of curves in Figure~\ref{fig:abar}.  The solid, dashed, and dotted curves are for $z=6,\,9$, and $12$.  In the early or middle stages of reionization, $\bar{A}<A_u$ because only the voids are ionized.  In this regime, $\bar{A}$ increases toward higher redshifts, because voids are denser (relative to the mean) at those times.  In the late stages, on the other hand, the dense clumps become ionized, $\bar{A}>A_u$, and $\bar{A}$ increases toward lower redshifts.  Because we follow the volume-averaged $\bxio$, we do not enter this regime until $\bxio \approx 1$:  dense clumps fill only a small fraction of the universe even though they contain a substantial fraction of the mass (see Fig. 2 of MHR00).  Thus the question of whether reionization is most efficient at high or low redshifts depends on the $\bxio$ of interest.  At low redshift, the early stages proceed faster, because $A_u$ is smaller and voids are deeper, but the late stages are much slower because dense structures are significantly more common; overall, reaching $\bxio \approx 1$ is easier at early times.

The shortcoming of this prescription is that it does not include the large-scale bias of ionized regions or the full inhomogeneity of the reionization process, in which some regions become highly ionized long before most of the universe.  In our model, the global recombination rate is simply a weighted average of the recombination rate inside each bubble:
\bq
\bar{A}_{\rm bub} = \int \deriv m_b \ A(m_b) \ \frac{m_b}{\bar{\rho}} \  n_b(m_b) \ [1 + \delta_x(z)].
\label{eq:abub}
\eq
Here the last factor is the mean physical overdensity of each bubble.  We set the recombination rate of each bubble, $A(m_b)$, by equating $\lambda_i$ and $R_b$ in order to fix $\Delta_i$.  

In Figure~\ref{fig:abar}, the upper thick set of curves shows $\bar{A}_{\rm bub}$ for the FZH04 model of equation (\ref{eq:dndm}); the redshifts are the same as in the lower set.\footnote{Note that we are being somewhat cavalier with the definition of ``$\bxio$'' here; in the FZH04 case it is the bubble filling factor.  The true ionized fraction (by volume) is slightly smaller because of the dense neutral clumps that fill a small volume of each bubble.}  This model increases $\bar{A}$ over that for uniformly distributed sources by a factor of about four, even early in reionization.  One reason is the biased location of bubbles, which sit in overdense regions.  Another is that the mean free path (and hence $\Delta_i$) grows much faster in big bubbles early on.  Because the sources are not distributed uniformly, they leave some low-density voids neutral while ionizing some higher-density gas, inevitably increasing $\bar{A}_{\rm bub}$.  The dot-dashed curve illustrates the relative importance of these effects:  it includes the variable mean free path but ignores the bubble overdensities.  In that case, $\bar{A}_{\rm bub}$ is significantly closer to the MHR00 value early in reionization, when $\delta_x(z)$ is largest, but approaches equation (\ref{eq:abub}) for $\bxio \rightarrow 1$ because $\delta_x \rightarrow 0$ in that limit.  According to Figure~\ref{fig:abar}, the two effects conspire to boost the recombination rate by a factor of a few throughout reionization.  
This again illustrates how recombinations ``tax" sources in high-density regions more heavily than those in low-density environments by forcing clusters of sources to blow large bubbles -- with correspondingly large recombination rates -- even early in reionization.

We must also note two caveats about this treatment.  First, we have ignored the recombination limit on the bubble size, and hence we overestimate $\bar{A}_{\rm bub}$ when $R_{\rm char} > R_{\rm max}$.  In actuality the enhancement will be smaller when $\bxio \ga 0.9$, but the degree will depend on the precise value of $R_{\rm max}$.  Second, if equation (\ref{eq:mfp-mhr}) overestimates the mean free path, $\bar{A}$ will increase by a larger factor because $\Delta_i$ rises.  For example, decreasing $\lambda_i$ by a factor of two increases $\bar{A}$ by $\sim 10$--$50\%$, depending on the ionized fraction and redshift.  

\section{An Observational Probe:  Lyman-series Absorption}
\label{obs}

One consequence of large ionized bubbles  is to reduce or eliminate the \lya ``damping wing" absorption by the IGM \citep{furl04-lya}.  If bubbles remained small throughout reionization, the \lya forest would allow \emph{no} transmission because the damping wings of the remaining neutral patches would overlap, even though \hii regions are still highly ionized \citep{barkana02-bub}.  With large bubbles, however, the neutral gas can lie sufficiently far away to render the damping wing negligible, opening a new window for studying the high-redshift universe:  quasar spectra should show interesting features even when $\bxio \sim 0.1$--$0.25$ (see, e.g., Fig.~9 of \citealt{furl04-lya}).

Our model shows that recombinations strongly affect these late stages, so the transmission in this regime constrains how the bubbles saturate once the IGM density structure begins to dominate.  To illustrate the power of this approach, we now extend the treatment of \citet{furl04-lya} to include the recombination limit.  We will ignore damping wing absorption in the following, because it is sub-dominant during the epoch of interest.  Given a model for the bubble topology and $P_V(\Delta)$, we can compute the probability that any patch of the IGM has an optical depth for transition $i$ smaller than $\tau_i$ by:
\bq
P(<\tau_i) = \int \deriv m_b \ n_b(m_b) \ \frac{m_b}{\bar{\rho}} \int_0^{\Delta_{\rm max}} \deriv \Delta \, P_V(\Delta),
\label{eq:ptau}
\eq
where $\Delta_{\rm max}$ is the maximum density for which $\tau<\tau_i$.  We assume ionization equilibrium within each bubble, so the neutral fraction is
\bq
x_{\rm HI} = \frac{\chi_e \bar{n}_H \alpha(T)}{\Gamma} \Delta,
\label{eq:ioneq}
\eq
where $\chi_e$ is a correction for (singly-ionized) helium.  
In the Appendix, we show that this assumption breaks down for gas with $\Delta > \Delta_i$, so we always take such regions to be fully neutral.  We assume an equation of state $T=T_0 \Delta^\gamma$, with $T_0=10^4 \kel$, and we approximate $\alpha \propto T^{-0.7}$. The ionizing rate per hydrogen atom is
\bq
\Gamma = \int \deriv \nu \  \epsilon_\nu \ \lambda_i \ \sigma_\nu.
\label{eq:gammadef}
\eq
Here the ionization cross-section is approximately $\sigma_\nu = \sigma_0 (\nu/\nu_0)^{-3}$, with $\sigma_0=6.3 \times 10^{-18} \cmsq$ and $h \nu_0=13.6 \eV$.  We will write the photon emissivity (by number) per unit frequency as $\epsilon_\nu = (\epsilon_0/\nu_0) \ (\nu/\nu_0)^{-\eta-1}$; the total emissivity is then $\epsilon_T=\epsilon_0/\eta$.  We take $\eta=3/2$ to approximate a starburst spectrum.  Finally, before reionization, the mean free path is $\lambda_i={\rm min}(R_b,R_{\rm max})$; it is frequency-independent in the limit that the bubbles are separated by walls of neutral hydrogen.  (We take a single $R_{\rm max}$ to represent the recombination limit at any instant, which appears to be a good assumption unless recombinations are quite rapid.)  Thus (c.f., \citealt{miralda03})
\bq
\Gamma = \frac{\lambda_i \sigma_0 \epsilon_T \eta}{3+\eta}.
\label{eq:gammares}
\eq
We can relate the emissivity to the efficiency parameter of FZH04 by setting $\epsilon_T=\zeta \bar{n}_b \chi_e^{-1} (\deriv f_{\rm coll,g}/\deriv t)$.  This assumes $\epsilon_i$ to be uniform throughout the universe; actually it is slightly larger around clusters of sources, but the fluctuations are small at the end of reionization.  We also note that $\zeta$ is here the instantaneous emissivity and should not include any factor accounting for past recombinations.  Finally, we take the local density field $P_V(\Delta)$ to be uncorrelated with the bubble topology; when $\bxio \approx 1$, this is probably not a bad approximation, though it must eventually be compared to high-resolution simulations.

\begin{figure}
\begin{center}
\resizebox{8cm}{!}{\includegraphics{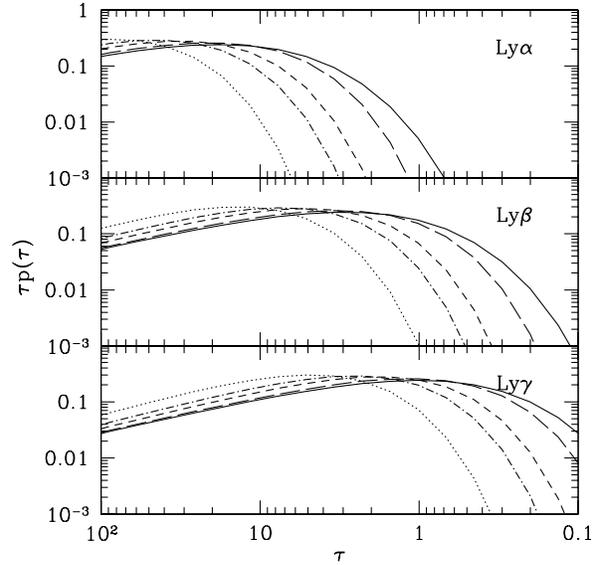}}\\%
\end{center}
\caption{Probability distribution of optical depth for the three Lyman transitions, if the universe has $\bar{x}_i=0.95$ at $z=6$.  The different curves assume $R_{\rm max}=10,\,20,\,30,\,60$, and $600$ Mpc (from left to right).}
\label{fig:ptx95}
\end{figure}

With these assumptions, we can write the local density as a function of observed optical depth,
\bqa
\Delta(\tau_i) & = & \left\{  170 \ \frac{\eta}{3+\eta} \ \left[ \frac{\alpha_A(10^4 \kel)}{\alpha(T_0)} \right] \ h(z) \ \left( \frac{\lambda_i}{{\rm Mpc}} \right) \right. \nonumber \\
& & \left. \times \ \zeta \ \left| \frac{\deriv f_{\rm coll}}{\deriv z} \right| \left( \frac{\tau_i}{\tau_{{\rm GP},i}} \right) \right \}^{1/(2-0.7 \gamma)},
\label{eq:deltatau}
\eqa
where $\tau_{{\rm GP},i}$ is the \citet{gunn65} optical depth for a fully neutral medium in transition $i$; substituting for $\Delta_{\rm max}$ in equation (\ref{eq:ptau}), we can then find the probability that any region of the IGM will have a small optical depth in our inhomogeneous bubble model.  

Figure~\ref{fig:ptx95} shows the resulting probability distributions for the Ly$\alpha$, Ly$\beta$, and Ly$\gamma$ transitions.  In each panel, the dotted, dot-dashed, short-dashed, long-dashed, and solid curves take $R_{\rm max}=10,\,20,\,30,\,60$, and $600 \Mpc$ (note that the last is essentially equivalent to setting $R_{\rm max}=\infty$).  We set $\bxio=0.95$ at $z=6$.  The distributions peak at a characteristic optical depth determined by the convolution of the bubble size distribution with the IGM density field.  Note how the peak moves relatively little for large $R_{\rm max}$, because it can never exceed the characteristic bubble size of FZH04.  While strong \lya transmission is always relatively rare, the Figure shows a sensitive dependence on $R_{\rm max}$:  strong transmission relies on deep voids, which are rare in the MHR00 model.  Thus moving the peak of $p(\tau)$ effectively moves the cutoff in the optical depth distribution and sharply reduces the leftover flux.  Similar, though somewhat smaller, differences occur for the other Lyman-series transitions.  The effects are less dramatic here because the higher transitions sample more representative parts of the density distribution \citep{oh05}.

\begin{figure}
\begin{center}
\resizebox{8cm}{!}{\includegraphics{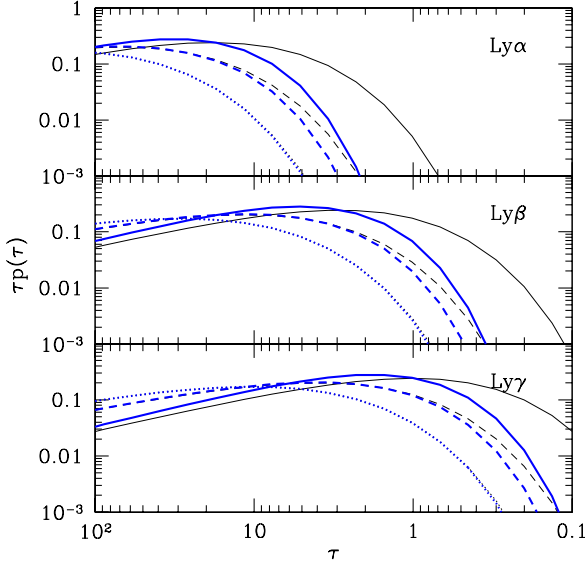}}\\%
\end{center}
\caption{Probability distribution of optical depth for the three Lyman transitions.  Within each panel, the thick curves assume $R_{\rm max}=30$ Mpc and the thin curves assume $R_{\rm max}=600$ Mpc.  The dotted, dashed, and solid curves are for $\bar{x}_i=0.75,\,0.85$, and $0.95$, respectively (all at $z=6$).}
\label{fig:ptzevol}
\end{figure}

Figure~\ref{fig:ptzevol} shows how the absorption varies with $\bxio$.  We assume $z=6$; the thick curves  take $R_{\rm max}=30 \Mpc$ while the thin curves take $R_{\rm max}=600 \Mpc$.  The dotted, dashed, and solid curves have $\bxio=0.75,\,0.85$, and $0.95$.  In the early stages, $R_{\rm max}$ makes little difference because $R_{\rm char} \ll R_{\rm max}$; hence, the dotted curves are coincident in each panel.  However, as the bubbles grow beyond $R_{\rm max}$, more and more of the universe becomes recombination-limited and thick and thin curves diverge (see also Fig.~\ref{fig:qrec}). At first only the tails of the distributions differ because $R_{\rm max} > R_{\rm char}$ at $\bxio=0.85$, but by the end of reionization even the peaks separate.  Again, the contrast is largest for \lya because that samples the most distant tail of the IGM density distribution.

Thus, the recombination limit ultimately determines the extent to which we can use quasar (or gamma-ray burst) spectra to study the late stages of reionization.  If $R_{\rm max}$ is small (less than 10 Mpc, for example), the pre-overlap era will have little or no transmission because the local ionizing background is small everywhere even if the damping wing is suppressed.  Hence the Gunn-Peterson trough would be complete.  If, on the other hand, $R_{\rm max}$ is comparable to the mean free path at $z \la 6$ ($\sim 20 \Mpc$; \citealt{fan02}), we expect non-negligible transmission as far back as $\bxio \sim 0.8$.  
Our model therefore makes several testable predictions for the influence of recombinations on observations. The key is that the scatter in apparent overlap increases with $R_{\rm max}$ while the allowed transmission over that range decreases.  The mean free path after overlap $\lambda(z\sim 6) \sim 20$ Mpc likely sets an upper bound on $R_{\rm max}$, because it normally increases monotonically with time. On the other hand, the significant Lyman series transmission along the line of sight to a $z \approx 6.4$ quasar \citep{white03}, coupled with the saturated absorption to a $z \approx 6.3$ quasar \citep{becker01}, sets a lower bound on $R_{\rm max}$. The observed scatter in the ``overlap redshift" $\Delta z$ should increase as $R_{\rm max}$ decreases.  Figure~\ref{fig:qrec} shows that taking $R_{\rm max}(z\approx 6$--$6.4) \sim \lambda(z\sim 6) \sim 20$ Mpc is consistent with the expected scatter in our model.  Moreover, $R_{\rm max}$ will not evolve strongly over that time (see Fig.~\ref{fig:pmfp}\emph{b}), barring strong  and sudden evolution in substructure -- e.g., photoevaporation of minihalos.  These arguments can be tightened both by observations of more high-redshift lines of sight and by more detailed simulations of the \lya forest at the end of reionization.
In general, the rate at which the transmission evolves will tell us how the universe passes from the ``bubble-dominated" topology characteristic of reionization to the smoother ``web-dominated" topology characteristic of the lower-redshift universe.  While 21 cm observations will teach us a great deal about the early and middle stages of reionization (when the 21 cm signal has a high contrast; FZH04, \citealt{furl04-21cmtop}) quasar spectra will likely remain our best probe of the tail end of reionization and this crucial overlap epoch. 

\section{Discussion}
\label{disc}

In this paper, we have studied how \hii regions evolve throughout reionization.  Building on the FZH04 model, in which ionized bubbles surround clusters of luminous sources, we first described how these bubbles combine and merge in the early and middle stages of reionization.  We derived bubble progenitor distributions and (approximate) merger rates analogous to the ``extended Press-Schechter" quantities commonly used for dark matter halos.  The distributions emphasize the ``inside-out" nature of reionization.  Large bubbles surrounding overdense regions begin ionizing earlier than average, while small bubbles (characteristic of voids) ionize later on.  They also show that \hii regions grow primarily through mergers with bubbles slightly larger than the current characteristic size.  
As a consequence, on a \emph{local} level reionization is more analogous to ``punctuated equilibrium" than ``gradualist" evolution, even if $\bxio$ increases smoothly.  Figure~\ref{fig:rw} shows several examples in which bubbles sit quietly for long periods between rapid growth spurts at major mergers. 

This behavior changes as we approach the end of reionization.  As the bubbles expand, internal recombinations consume an ever growing fraction of ionizing photons.  In other words, once a bubble's size becomes comparable to the mean free path of an ionizing photon, it stops growing and becomes a true ``Str{\" o}mgren sphere."  We showed how this saturation point can be incorporated into the FZH04 model by building on the picture presented by MHR00.  Note that once a significant fraction of the universe sits within such bubbles, the bubble sizes do not meaningfully reflect variations in the mean free path of ionizing photons. In this regime, our model describes large-scale fluctuations in the mean free path.  This makes interpreting observations somewhat subtle:  direct measures of the ionized volumes, such as 21 cm studies of the high-redshift universe (e.g., \citealt{madau97,ciardi03-21cm,zald04,furl04-21cmtop}), will see large contiguous ionized regions but can tell us little about how the mean free path evolves and hence how reionization actually ends.  Instead, observations of Lyman-series absorption toward high-redshift quasars and of Ly$\alpha$-emitting galaxies may remain our best probes of the overlap era, because those techniques are much more sensitive to the strength of the ionizing background and hence to the mean free path of ionizing photons. 

In IGM models similar to MHR00, bubbles saturate only on scales $\ga 20$ comoving Mpc.  However, applying the MHR00 model at $z \ge 6$ requires extrapolation from their simulation fits.  In particular, the simulations assumed that the IGM is smooth on the Jeans scale of photoionized gas ($T \sim 10^4 \kel$).  This obviously need not be the case for neutral gas at high redshifts.  In principle the gas may be colder than the CMB and a large fraction could condense into minihalos, which act as sinks of ionizing photons.  Minihalos will significantly reduce the mean free path and hence freeze the bubble growth earlier.  We considered an extremely simple model to illustrate their possible effects; a more detailed look must consider the photoevaporation of these objects as they are ionized \citep{barkana99,shapiro04}, which makes the problem time-dependent.  Nevertheless, our picture fits with the current observational constraints.   \citet{fan02} measured the mean free path of ionizing photons at $z \sim 5.5$ to be $\sim 20 \Mpc$ (using the MHR00 model).  If the MHR00 model is accurate at high redshifts, then the bubbles grow smoothly until this limit is reached when $\bxio \ga 0.9$ (see Fig.~\ref{fig:qrec}).  If the IGM is significantly more clumpy, the universe may pass through a long highly-ionized phase in which the mean free path grows slowly as, e.g., minihalos photoevaporate.  The true picture awaits a better understanding of the role of feedback and IGM heating in suppressing small-scale structure in the IGM \citep{oh03-entropy}.

Another effect of our description of inhomogeneous reionization is to increase the recombination rate of the universe above that predicted by MHR00 (by up to a factor $\sim 4$), especially when $\bxio$ is relatively small.  This occurs because ionized bubbles originate around clusters of galaxies (which in turn occur in overdense regions) and because, within each bubble, the minimum density to which the IGM must be ionized is larger than for a uniformly illuminated universe.  
Thus we have found that recombinations exact a ``progressive tax" on the ionizing sources that becomes larger around the largest proto-clusters of sources:  any such group that can blow a large bubble into the IGM will soon have its ionizing photons consumed by recombinations, although its more isolated neighbors remain effective ionizing agents. This helps to explain the simulations of \citet{ciardi03-sim}, who argued that the densest regions are the first to begin reionization but the last to ``complete" it.
 
Most importantly, our model shows that the radiation background at any given point in the IGM will evolve through a series of discrete jumps until it saturates at the mean free path.  We must therefore be careful in interpreting jumps in the ionizing background (or in the IGM transmission) -- every point will experience a sequence of jumps throughout reionization, with the final (and likely largest) jump occurring when a region becomes recombination-limited.  After this point the ionizing background evolves only slowly (see Fig.~\ref{fig:rw}).  Although most regions reach this stage over a relatively short time interval (see Fig.~\ref{fig:qrec}), some bubbles grow rapidly and saturate at much earlier times, particularly if the limit appears when $R\la 10$ comoving Mpc.  Once a bubble reaches this point, reionization is complete for all intents and purposes within that volume, regardless of the true global $\bxio$ (c.f., \citealt{gnedin00}), and it is obvious that we can no longer treat the ``overlap epoch" as a single global event.  Instead these large bubbles enter a ``post-overlap" phase when $\bxio <1$!  As a result, we expect substantial cosmic variance in the \emph{apparent} onset of reionization along different lines of sight.  Note that this is a qualitatively different source of variance than those described by \citet{wyithe04-var}, who attributed differences in the onset of reionization to a combination of the light-travel time across bubbles and modulation by the underlying density field (similar to the pure FZH04 model).  While the latter is crucial to the large-scale topology of reionization, the end stages of reionization are actually driven by fluctuations in the mean free path within large contiguous \hii regions and not simply by source clustering.  In comparison to their model, we predict larger scatter in the ``overlap" time but smaller scatter in the mean free path during that interval, because every region's size must be near the Str{\" o}mgren limit.

This limit $R_{\rm max}$ determines the ionizing background within the bubbles (and hence the transmission, given a density field and equation of state).  Thus observations of Lyman-series transmission in quasar (or gamma-ray burst) spectra can constrain how substructure in the IGM assumes control of the metagalactic radiation field and confines the bubbles.  If, for example, minihalos limited the mean free path to Mpc scales, significant transmission cannot occur until reionization is truly complete; on the other hand, density models similar to MHR00 predict measurable transmission even when $\bxio \sim 0.75$.

Our model also provides a clear and intuitive interpretation of the differences between the two best-studied lines of sight to $z>6$ quasars \citep{becker01,white03}, which go much deeper than differences in the endpoint of the Gunn-Peterson trough.  One shows apparent complete absorption, while the other remains highly ionized and has a number of transmission features at $z >6$.  These differences are intriguing because they appear to be coherent over $\sim 100$ comoving Mpc, on which scales density fluctuations should be miniscule at such high redshifts.  In our model, the difference arises because overlap occurs somewhat earlier along one line of sight.  The FZH04 model shows that reionization effectively amplifies small density fluctuations by distributing the IGM gas through a range of bubble sizes.  Here we have shown that, during the end stages of reionization, substructure in the IGM further modulates the mean free path and hence the expected transmission.  Thus, at least qualitatively, it appears reasonable that the line of sight to the $z=6.28$ quasar may pass through a slightly underdense region along which the \hii regions have not yet reached the Str{\" o}mgren radius, suppressing the transmission relative to the other quasar.  In any  case, we have shown that such large-scale fluctuations are a generic feature of reionization and can at least in principle be used to infer some key properties of the IGM.

All that said, our model is approximate and must be improved through comparisons to observations.  FZH04 noted several simplifying assumptions; these included the inhomogeneous distribution of sources inside bubbles, the possibility that $\zeta$ varies across galaxies, and the Lagrangian treatment of purely linear fluctuations.  However, the most important were recombinations and purely spherical bubbles.  We have shown that recombinations do affect the bubble growth, but only in the late stages of reionization, helping to validate the FZH04 model.  However, all of our conclusions are subject to caveats about the unknown density distribution within the IGM.  That is something that clearly must be simulated more carefully in the future if we are to understand the full effects of recombinations.  It promises to be difficult -- following the growth of small-scale structure, including the feedback of reionization on it (including the photoevaporation of minihalos; \citealt{shapiro04}), while also sampling a representative volume of the universe ($\ga 100 \Mpc$) -- will tax even the promise of Moore's law.  A hybrid strategy of bootstrapping small boxes, which resolve the IGM structure, into larger, lower-resolution boxes may be the most viable.

The other crucial assumption is that of spherical bubbles.  We argued in \S \ref{fzh} that this will probably break down when the bubbles fill a large fraction of the universe, because then ionized gas can surround isolated islands of neutral gas.  In this case the bubble size may describe the typical separation of such islands.  This too must be tested in simulations, or at least in numerical realizations of our barrier formalism (e.g., \citealt{zahn05}).  However, our recombination-based model is less subject to such problems.  We have seen that, once the bubbles become large, they fragment into Str{\" o}mgren spheres; our model describes the distribution of mean free paths.  This quantity is not so sensitive to the underlying topology of ionized and neutral gas, because it supposes only the existence of large continuous regions of ionized material.

\vspace{0.1cm}

We thank M.~McQuinn for useful discussions. We are grateful to the KITP for hospitality during the beginning of the project, under NSF grant PHY99-07949. SPO also acknowledges NSF grant AST-0407084 for support. 


\appendix

\section{A Closer Look at the MHR00 Model}

The interior of a Str{\" o}mgren sphere during reionization is similar to the universe at a lower redshift, when the Ly$\alpha$ forest is in ionization equilibrium. Our physical picture of this situation is heavily influenced by numerical simulations, which can reproduce the $z \sim 3$ observations remarkably well by assuming a uniform radiation field $\Gamma$. In this regime, $x_{\rm HI} \propto \Delta$. At face value, this is rather different from the MHR00 assumption that $x_{\rm HI} \approx 0$ for $\Delta < \Delta_{i}$ and $x_{\rm HI} \approx 1$ for $\Delta > \Delta_{i}$. Here, we explore whether these two views are compatible. In particular, we compute the column density distribution function $f(N_{\rm HI})$ and compare the resulting mean free path with the MHR00 ansatz of equation (\ref{eq:mfp-mhr}).  We  also check the validity of the MHR00 assumption that absorption is dominated by Lyman limit systems rather than the cumulative photoelectric opacity of lower column density systems. As a byproduct, we derive a number of useful scaling relations. Throughout this Appendix we only consider recombination-limited bubbles, where $\epsilon \approx A$. 

It is useful to develop our physical intuition by making some (surprisingly accurate) analytic approximations to the MHR00 model and examining the salient scalings.  Let us consider spherical absorbers with density profiles $\rho \propto r^{-\alpha}$. Suppose a radiation field $\Gamma$ is incident upon the absorbers.  We define a Str{\" o}mgren surface via $\Gamma r^{2} \propto \rho^{2} r^{3}$, or $r \propto \Gamma^{1/(1-2\alpha)}$. Since $r \propto \Delta^{-1/\alpha}$, this implies a relation between the radiation field and the overdensity up to which the absorber is ionized, $\Gamma \propto \Delta^{(2\alpha-1)/\alpha}$. The mean free path of ionizing photons is inversely proportional to the cross-section of the absorber, $\lambda \propto 1/r^{2} \propto \Delta^{2/\alpha}$. Finally, the recombination rate $A \propto \rho^{2} r^{3} \propto r^{-2\alpha+3} \propto \Delta^{(2\alpha-3)/\alpha}$. For an isothermal sphere, $\alpha=2$, these relations imply that $\Gamma \propto \Delta^{3/2}$, $\lambda \propto \Delta$, and $A \propto \Delta^{1/2}$. Note that self-consistently, $\Gamma \propto \epsilon \lambda \approx A \lambda \propto \Delta^{3/2}$. 

\begin{figure*}
\begin{center}
\resizebox{16.13cm}{!}{\includegraphics{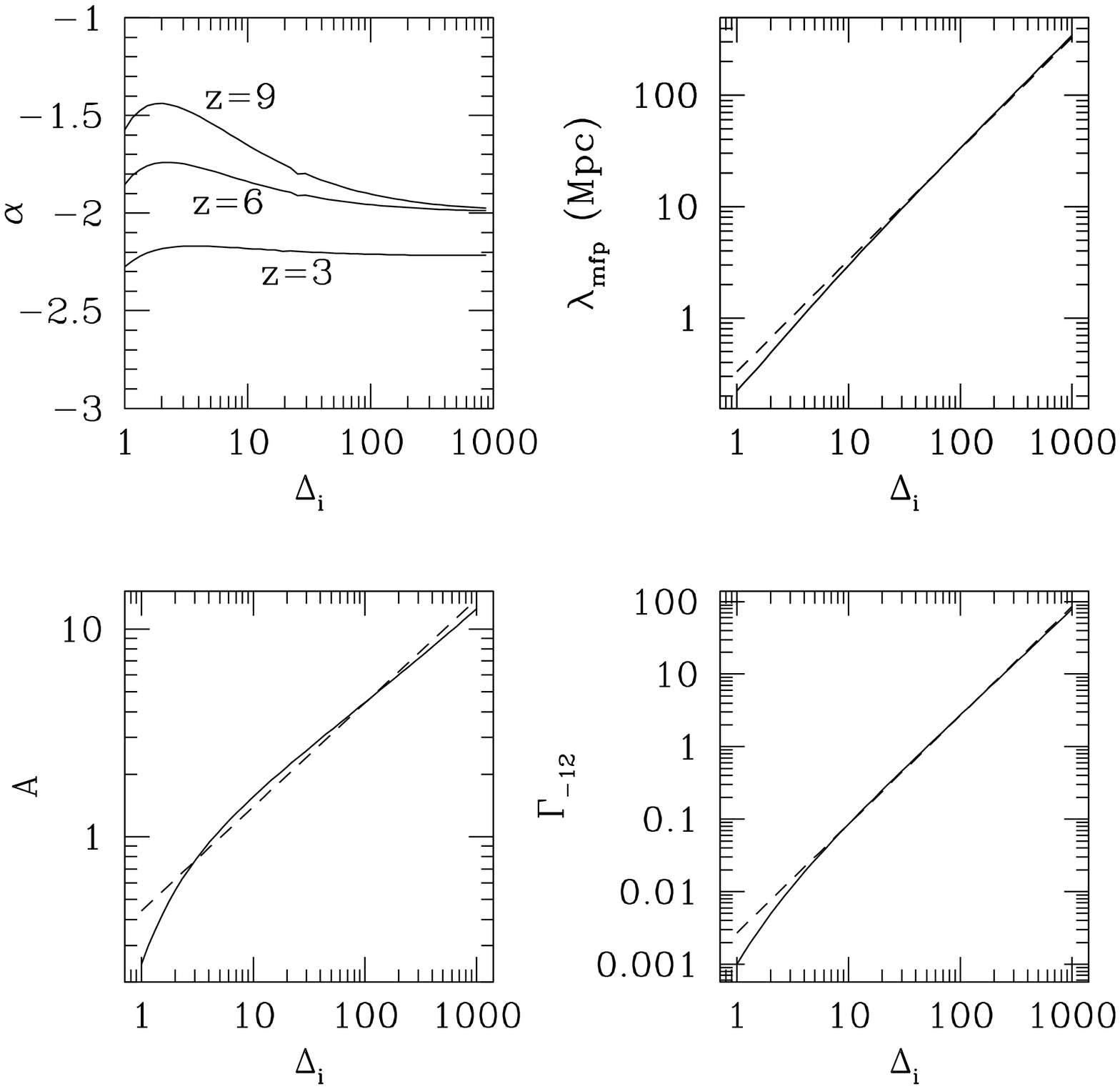}}\\%
\end{center}
\caption{\emph{Top left}: The logarithmic slope $\alpha$ of the density profile, from equation (\ref{eq:density_slope}). The density profiles quickly reach their asymptotic power law scalings. The other panels show the proper mean free path $\lambda_{\rm mfp}$, recombination rate $A$, and ionization rate $\Gamma_{-12}$, as computed numerically (solid lines) and via the power law scaling relations (equation \ref{eq:scaling_relations}), at $z=6$. Comparison at other redshifts show similarly good agreement.}
\label{fig:MHR_scalings}
\end{figure*}

By design, the MHR00 density distribution function, equation (\ref{eq:pvd}), asymptotes to a power law density profile $\Delta \propto r^{-3/(\beta-1)}$ for $\Delta \gg 1$. At $z=3$, $\beta=2.35$ (derived from a fit to numerical simulations), corresponding to $\rho \propto r^{-2.22}$, while for $z \ge 6$, MHR00 assume isothermal density profiles $\rho \propto r^{-2}$, implying $\beta=2.5$. How quickly does the MHR00 density distribution reach this limiting behaviour?  In their treatment, the radius of an absorber is $r \propto [1-F_{V}(\Delta_{i})]^{1/3}$. Hence, we can compute the effective slope of the density power law numerically via
\begin{equation}
\alpha = -\frac{{\rm d(log} \Delta)}{{\rm d(log}r)} = -\left[ \frac{{\rm d(log} (1-F_{V})^{1/3})}{{\rm d(log}\Delta)}\right]^{-1}, 
\label{eq:density_slope}
\end{equation}
which we show in the upper left panel of Figure~\ref{fig:MHR_scalings} for $z=(3,6,9)$. The density distribution quickly approaches the limiting power law density profiles for the overdensities of interest. Thus the approximations for isothermal density profiles should be accurate for the regime of interest. Because the number density of absorbers evolves with redshift (depending on the details of structure formation), the normalization has to be obtained numerically.  We find for $z=(3,6,9)$:
\begin{eqnarray}
\lambda_{\rm mfp} &=& (41, 33, 33) \left( \frac{\Delta_{i}}{100} \right) \ {\rm Mpc} \nonumber \\
A  &=& (6.8,4.4,3.0) \left( \frac{\Delta_{i}}{100} \right)^{1/2} \nonumber \\
\Gamma_{-12} &=& (0.42,2.7, 9.1) \left( \frac{\Delta_{i}}{100} \right)^{3/2}
\label{eq:scaling_relations}
\end{eqnarray}
where $\lambda_{\rm mfp}$ is in physical (proper) units, $A$ is in units of recombinations per baryon per Hubble time, $\Gamma_{-12} \equiv \Gamma/(10^{-12} \, {\rm s^{-1}})$, and we have used equation (\ref{eq:gammadef}) to evaluate $\Gamma$. These power law scaling relations are compared to the exact expressions (solid lines) in Figure~\ref{fig:MHR_scalings} at $z=6$.  They are remarkably accurate and useful for quick estimates. 

The MHR00 model postulates that $x_{\rm HI} \approx 0$ for $\Delta < \Delta_{i}$ and $x_{\rm HI} \approx 1$ for $\Delta > \Delta_{i}$; this is rather different from the usual ionization equilibrium assumption ($x_{\rm HI} \propto \Delta$) in \lya forest studies.  MHR00 argued that their model could accommodate such a scaling if one defined $\Delta_{i}$ such that $x_{\rm HI}(\Delta_{i})=0.5$.  Then, assuming ionization equilibrium, one obtains virtually the same recombination rate $A(\Delta_{i})$ as with the original definition of $\Delta_{i}$ (see their Figure 2). However, it predicts the wrong scaling between $\Delta_{i}$ and the radiation field: if $x_{\rm HI}(\Delta_{i})=$const, then $\Delta_{i} \propto \Gamma$. By contrast, equation (\ref{eq:scaling_relations}) implies that $\Delta_{i} \propto \Gamma^{2/3}$. Indeed, from the scaling relations in equation (\ref{eq:scaling_relations}), we have
\bqa
x_{\rm HI}(\Delta_{i}) & \approx & \frac{\chi_{e}\bar{n}_{\rm H} \alpha(T)}{\Gamma(\Delta_{i})}\Delta_{i} \\
& = & (1.4, 1.2,1.0) \times 10^{-3} \left( \frac{\Delta_{i}}{100} \right)^{-1/2}, \nonumber
\label{eqn:xHI}
\eqa
apparently contradicting the MHR00 assumption that $x_{\rm HI}(\Delta_{i}) \sim O(1)$. What has gone wrong?

The key is that we must take the finite absorber sizes -- and in particular self-shielding -- into account.  We therefore need to associate a column density $N_{\rm HI}$ with a given overdensity $\Delta_{i}$.  \citet{schaye01} has shown that many observed properties of the Ly$\alpha$ forest can be easily understood if the typical size of an overdensity $\Delta$ is the local Jeans length $L_{J}$.   Then, assuming $N_{\rm HI} \approx x_{\rm HI} \Delta \bar{n}_{H} L_{J}$ and photo-ionization equilibrium, we obtain
\begin{equation}
N_{\rm HI} = 3.3 \times 10^{17} \, {\rm cm^{-2}} \left( \frac{\Delta}{100} \right)^{3/2} T_{4}^{-0.26} \Gamma_{-12}^{-1} \left( \frac{1+z}{7} \right)^{9/2}.
\label{eqn:NHI}
\end{equation}
A Lyman-limit system which can self-shield from the surrounding radiation field has $N_{\rm HI} \approx 1/\sigma_{0} = 1.6 \times 10^{17} \, {\rm cm^{-2}}$.  From equation (\ref{eqn:NHI}), an object enters this regime at
\begin{equation}
\Delta_{\rm LLS} = (330, 61, 21) \ \Gamma_{-12}^{2/3}
\label{eqn:delta_LLS}
\end{equation}
for $z=(3,6,9)$. Note that this expression now has the correct power-law scaling. On the other hand, from equation (\ref{eq:scaling_relations}), we have
\begin{equation}
\Delta_{i} = (180,51,23) \ \Gamma_{-12}^{2/3}.
\label{eqn:delta_i}
\end{equation}
Thus, $\Delta_{i} \approx \Delta_{\rm LLS}$. From equations (\ref{eqn:xHI}), (\ref{eqn:delta_LLS}), and (\ref{eqn:delta_i}), we conclude that the MHR00 model is indeed fully self-consistent, in the sense that $x_{\rm HI} \approx 0$ for $\Delta < \Delta_{i}$, and $x_{\rm HI} \approx 1$ for $\Delta > \Delta_{i}$. 
This happens because MHR00 calibrated their model to numerical simulations, which obviously do contain the relevant physical scales for the absorbers.  There is an abrupt transition at $\Delta_{i}$ due to self-shielding, so that $\Gamma \rightarrow 0$ for $\Delta > \Delta_{i}$. Below the threshold, ionization equilibrium with $\Gamma \approx$ const and $x_{\rm HI} \propto \Delta$ are accurate.

Equation (\ref{eqn:NHI}) also allows us to calculate more detailed statistics of the Ly$\alpha$ forest and to compare with observations at $z\sim 3$.  In particular, let us calculate the canonical distribution function,
\begin{equation}
f(N_{\rm HI},z)\equiv  \frac{\deriv^{2}N}{\deriv N_{\rm HI} \deriv z} \frac{H(z)}{H_{0}} \frac{1}{(1+z)^{2}}. 
\end{equation}
Observe that:
\bqa
\Omega_{g}& = & \Omega_{g} \int \deriv \Delta P_{V}(\Delta,z) \Delta \nonumber \\
& = & \frac{8 \pi G m_{\rm H}}{3 H_{0} c (1-Y)} \int N_{\rm HI} x_{\rm HI}^{-1} f(N_{\rm HI},z) \deriv N_{\rm HI}.
\eqa
Hence, differentiating both sides, we obtain
\begin{equation}
f(N_{\rm HI},z)=\Omega_{b}\frac{\deriv \Delta}{\deriv N_{\rm HI}} \Delta P_{V}(\Delta,z) \frac{3 H_{0} c (1-Y)}{8 \pi G m_{\rm H}} x_{\rm HI} N_{\rm HI}^{-1}.
\label{eqn:fN}
\end{equation}
Using equation (\ref{eqn:NHI}) and our power-law scalings for $P_{V}(\Delta)$, we have, at $z=(3,6,9)$,
\bqa
f(N_{\rm HI}) & = & (4.5,4.2,4.7) \times 10^{-18} \ N_{17}^{(-1.56,5/3,5/3)} \nonumber \\ 
& & \times \ \Gamma_{-12}^{-(0.56,2/3,2/3)},
\label{eqn:fN2}
\eqa
where $N_{17} \equiv N_{\rm HI}/(10^{17} \, {\rm cm^{-2}})$. The different power law exponents arise because we have assumed $\beta=2.35$ at $z=3$ but $\beta=2.5$ at $z\ge 6$. In general, $f(N_{\rm HI}) \propto \Delta^{-\beta} \Gamma \propto N_{\rm HI}^{-2\beta/3} \Gamma^{-2\beta/3 +1}$. 

To test our model, we compare to observations of the Ly$\alpha$ forest at $z\sim3$. The unknown radiation field $\Gamma_{-12}$ can be fixed through matching the mean flux decrement in quasar spectra.  \citet{schirber03} find that $\Gamma_{-12}(z)=10^{-0.24z+0.4}$ is a good approximation to the measurements of \citet{mcdonald01} in the redshift range $z=2.3-5$. If we insert this radiation field into equation (\ref{eqn:fN}), we find excellent agreement with observations. For instance, \citet{hu95} find a best fit 
power law $f(N_{\rm HI})=5.5 \times 10^{-12} N_{13}^{-1.46}$ over the range $N_{\rm HI}=2 \times 10^{12}-3 \times 10^{14} \ {\rm cm^{-2}}$ at $z=2.8$, whereas equation (\ref{eqn:fN}) predicts $f(N_{\rm HI})=7.6 \times 10^{-12} N_{13}^{-1.56}$. Likewise, \citet{storrie94} find an abundance of Lyman-limit systems $dN_{\rm LL}/dz=3.3 \pm 0.6$ at $z=4$. By comparison, if we integrate equation (\ref{eqn:fN}) over all column densities $N_{\rm HI} > N_{\rm LLS} = 1.6 \times 10^{17} {\rm cm^{-2}}$, we find $\deriv N_{\rm LLS}/\deriv z=3.8$. Thus our model seems physically reasonable, although the observations suggest rescaling the normalization of equation (\ref{eqn:fN}) by a factor $f_{s}\sim 0.7$ (probably due to some ambiguity in the definition of $L_{J}$). As a further consistency check, equation (\ref{eqn:fN}) predicts a mean free path
\begin{equation}
\lambda_{\rm mfp} = \frac{c H^{-1}}{(1+z)} \left[ \frac{\deriv N}{\deriv z} \right]^{-1} = (76,20,6.4) \ \Gamma_{-12}^{2/3} \left( \frac{f_{s}}{0.7} \right) \ {\rm Mpc}.
\end{equation}
at $z=(3,6,9)$. By comparison, equation (\ref{eq:scaling_relations}) yields $\lambda_{\rm mfp}= (73,17,7.6) \ \Gamma_{-12}^{2/3}$Mpc, which is in good agreement. 

The sole remaining point to check is the relation between the mean free path $\lambda_{\rm mfp}$ of ionizing photons and the distance between Lyman-limit systems $\lambda_{\rm LLS}$; we have assumed that they are roughly comparable, $\lambda_{\rm mfp} \approx \lambda_{\rm LLS}$. This amounts to equating Lyman-limit systems with contours of overdensity $\Delta_{i}$ and neglecting the cumulative photoelectric opacity of lower column-density systems.  In general, if $f(N_{\rm HI}) \propto N_{\rm HI}^{-\delta}$, then the effective optical depth $\tau_{\rm eff} \propto N_{\rm HI}^{-\delta+2}$; as long as $\delta <2$, the highest column density systems dominate the effective optical depth and hence determine the mean free path. More precisely, the absorption probability per unit length, $\lambda_{0}^{-1}$, for a photon at the hydrogen ionization absorption edge is \citep{miralda03}:
\begin{equation}
\frac{1}{\lambda_{0}}= \frac{\int_{0}^{\infty} d\tau \, \tau^{-\delta} (1-e^{-\tau})}{\int_{1}^{\infty} d\tau \, \tau^{-\delta} \lambda_{\rm LLS}} = \frac{(2.0,2.6)}{\lambda_{\rm LLS}} 
\end{equation}
for $\delta=2\beta/3=(1.56,5/3)$ at $z=3$ and $z\ge 6$ respectively.  Here we have defined $\tau=N_{\rm HI}/(1.6 \times 10^{17} {\rm cm^{-2}})$. Thus, we see that for the assumed model the mean free path is indeed comparable to the spacing between Lyman-limit systems. 
At the Lyman edge, it is actually a factor $\sim 2$ smaller, but higher energy photons will reduce some of that difference.

Note that $\lambda_{\rm mfp}/\lambda_{\rm LLS}$ depends only on the slope of column density distribution function $\delta$, which in turn depends on the slope $\alpha$ of the density profile around absorbers. If they are shallower, photoelectric absorption from lower column-density systems becomes more important.  In such a picture, absorption is cumulative and continuous rather than like a picket-fence. In particular, for $\alpha<1.5$, $\delta>2$ and the opacity is dominated by low column density systems (at $\delta=2$ itself, we find $\lambda_{\rm mfp}/\lambda_{\rm LLS} \sim 0.1$). Note also that since $A \propto \Delta^{(2\alpha-3)/\alpha}$, recombinations are dominated by low overdensities rather than high overdensities for $\alpha < 1.5$. In this case, the recombination rate converges once emission exceeds a certain threshold, allowing the Str{\" o}mgren surface around ionized clumps to suddenly move inward and the mean free path to increase dramatically (see also the discussion in \citealt{madau99}). Thus, the nature of percolation itself depends quite sensitively on the unknown slope of the density field in dense clumps. One could imagine that gas density profiles might be quite different at high redshift, when potential wells are much shallower and Jeans smoothing more important. For instance, if there is a period of early reionization that raises the entropy of the IGM, gas density profiles around minihalos can be much shallower than the dark matter profiles \citep{oh03-entropy}. 

\end{document}